\documentclass[aps,prb,superbib,superscriptaddress,amsfonts,preprint]{revtex4}

\usepackage{graphicx,epsfig,tabularx}
\usepackage{amssymb,amsmath}
\usepackage{multirow,xcolor}
\usepackage{filecontents}
\usepackage{hyperref}
\hypersetup{
	colorlinks = true,
	linkcolor = blue,
	linkbordercolor = white,
	citecolor=blue,
}
\usepackage{subfig,caption}
\newcommand{\ket}[1]{\ensuremath{\left|#1\right\rangle}}

\newcommand{\avg}[1]{\ensuremath{\left\langle #1 \right\rangle}}
\newcommand{\bm}[1]{\boldsymbol{#1}}

\newcommand{\refeq}[1]{Eq. \ref{#1}}
\newcommand{\refsec}[1]{Sec. \ref{#1}}
\newcommand{\reffig}[1]{Fig. \ref{#1}}
\newcommand{\reftable}[1]{Table. \ref{#1}}

\begin{document}
\title{ Vibrational Relaxation at a Metal Surface: Electronic Friction Versus Classical Master Equations }
\author{Gaohan Miao}
\affiliation{Department of Chemistry, University of Pennsylvania, Philadelphia, Pennsylvania 19104, USA}
\author{Wenjie Dou}
\affiliation{Department of Chemistry, University of Pennsylvania, Philadelphia, Pennsylvania 19104, USA}
\author{Joseph Subotnik}
\email{subotnik@sas.upenn.edu}
\affiliation{Department of Chemistry, University of Pennsylvania, Philadelphia, Pennsylvania 19104, USA}

\begin{abstract}
	\label{abstract}
	Within a 2-D scattering model, we investigate the vibrational relaxation of an idealized molecule colliding with a metal surface. Two perturbative nonadiabatic dynamics schemes are compared: 
	$(i)$ electronic friction (EF)
	and $(ii)$ classical master equations (CME).
	In addition, we also study a third approach, $(iii)$ a broadened classical master equation (BCME) that interpolates between approaches $(i)$ and $(ii)$. Two conclusions emerge. First, even though we do not have exact data to compare against, we find there is strong evidence suggesting that EF results may be spurious for scattering problems with more than one nuclear dimension. Second, we find that there is an optimal molecule-metal coupling that maximizes vibrational relaxation rates by inducing large nonadiabatic interactions.
\end{abstract}

\maketitle

\section{Introduction}\label{sec: introduction}

	Non-adiabatic dynamics are known to play an essential role in photochemistry and excited state dynamics in the gas phase and in solution. 
	For such experiments\cite{Wong2002SolventDissipation, Wong2002SolventDissipation2, Jailaubekov2013PhotoExcitation, Nelson2014PhotoExcitationDynamicsRev, Schwartz1996CondensePhaseNonAdiab, Schwerdtfeger2014SolutionNonadiab}, it is obvious that photo-excitation is followed by energetic relaxation as electrons relax and nuclei heat up: after all, the density of states for nuclear motion is much larger than the density of states for electronic motion and thus, after a long time, all electronic energy must be converted into heat (or nuclear motion). Thus, in solution or gas phase photochemistry, nonadiabatic dynamics are paramount.

	Now, using the same logic, consider the role of nonadiabatic effects at a metal surface.
	On the one hand, a bulk metal carries an enormous density of phonon states; these states will usually be the final acceptors of any excess energy and thus act as a driving force for nonadiabatic transitions.
	On the other hand, a metal also carries a large density of electronic states, and thus electronic transitions are possible (even without nuclear motion). Thus, there is no guarantee that one will observe nonadiabatic dynamics near metal surfaces, i.e. a nontrivial coupling of nuclear motion with electronic transitions.
	In a series of recent papers, however, the Wodtke group has given very convincing evidence that nonadiabatic effects {\em are ubiquitous } when studying scattering processes for molecules off of metal surfaces. In the most famous experiments, the Wodtke group has scattered NO molecules across Au(111) surfaces\cite{Wodtke2003NOLiF,Huang2000NOAuexp,White2005NOAuExp_vibrelax,Wodtke2004nonadiabaticRev,Bartels2011nonadiab_minirev,Wodtke2016Rev,Bartels2014NOAu_moreexp}, and found clear evidence that vibrational relaxation is mediated by nonadiabatic processes (in this case, transient electron transfer). Thus, there is currently a great deal of interest in the physics and chemistry communities regarding how to model these difficult experiments.

	Now, obviously, with a metal substrate, a fully quantum description of the nuclear and electronic degrees of freedom would be prohibitively expensive.\cite{Katz2005ModelForNOAu}
 For realistic, multi-atom simulations of the Wodtke experiments, semiclassical treatments are the only possible way forward. And, in this context, there are two well known perturbative limits.\cite{Galperin2015MoleculeMetalInterface}
	$(i)$  On the one hand, for weak nonadiabatic effects  (i.e. strong metal-molecule couplings and weak electron-phonon couplings), the usual semiclassical framework is to assume that the molecular motion on the metal surface feels so-called ``electronic friction'' from the bath of metallic electrons. This concept of electronic friction has been used many times in the past\cite{HeadGordon1992COCutheory,Rittmeyer2015EF_application,Nitzan1983EF_compare_to_quantum_pertubation,Li1992EFsimu}, most famously by Head-Gordon and Tully to study the relaxation of CO on a Cu substrate\cite{HeadGordon1992COCutheory}.
	$(ii)$ On the other hand, for strong nonadiabatic effects (i.e. weak metal-molecule coupling and strong electron-phonon  couplings), another approach is a classical master equation, whereby molecules move as if they are charged or uncharged with stochastic hops between different charge states. This master equation approach describes processes known as dynamics induced by multiple electronic transitions (DIMET) by the surface-science community \cite{Frischkorn2006NonadiabaticReactionRev, Misewich1992DIMET}.
	In the context of standard electron transfer theory, one imagines that the former approach should describe inner-sphere heterogeneous electron transfer and the latter approach should describe outer-sphere heterogeneous electron transfer\cite{Bard2010HeterogenousElectrodeReaction}.

	Unfortunately, for the case of NO scattering off of gold, neither EF or CME may be valid.
	First, the NO-Au interaction is not small at short distances, and so the CME approach is likely inapplicable.  
	Second, far from the metal, the NO-Au interaction is weak and the Wodtke group has given three pieces of evidence suggesting that electronic friction is inapplicable for scattering problems\cite{Huang2000NOAuexp,White2005NOAuExp_vibrelax,Wodtke2003NOLiF}:

	\begin{enumerate}

		\item For NO incoming in a highly excited state (e.g., $n_{vib} = 15$), vibrational relaxation shows no barrier (as one would expect with adiabatic dynamics). Furthermore, the most probable exit channel has $n_{vib} = 7$. However, if the metal Au is replaced by an insulator, LiF, a barrier does appear and the most probable exit channel is the original vibrational state (and the second most probable exit channel is the original vibrational state minus one). This data obviously suggests strong coupling between electronic transitions in the metal and vibrational transitions in the molecule and the resulting dynamics clearly demonstrate sudden as opposed to gradual changes of state (not as we would expect with electronic friction).

	\item For NO incoming in a highly excited state (e.g., $n_{vib} = 15$), one can observe hot electron emission from the metal at very large kinetic energies. Such emission precludes simple electronic friction descriptions based on fast electronic equilibration.

	\item For NO incoming in the ground state, Wodtke {\em et al} have observed that multiple quanta can be excited directly, which would not agree with a frictional description, i.e. a golden-rule picture of the dynamics assuming small electron-phonon couplings.

	\end{enumerate}

	Thus, the Wodtke experiments present a clear challenge to theoretical chemistry and physics. Since the relevant dynamics have strong electron-phonon couplings, and because the metal-molecule couplings can be weak or strong (depending on the distance to the metal), and since transient electron transfer cannot be ignored, the Wodtke experiments simply do not sit in any simple perturbative regime.\cite{Galperin2015MoleculeMetalInterface} To model these difficult dynamics, a few years ago, Shenvi {\em et al} suggested discretizing the metal continuum and they developed the so-called independent-electron surface hopping (IESH) approach\cite{Shenvi2009IESH, Shenvi2009NOAusimu_sci}.
Quantum master equations have also been proposed for this purpose. \cite{Monturet2010NOAu_EF_simu,Li2002MCmodelforNOAu}.
	More recently, by rederiving the origins of electronic friction, our research group showed how to extrapolate between the CME and EF regimes, so that one could develop a universal, semiclassical nonadiabatic dynamics algorithm for strong or weak coupling near a  metal surface. We labeled the resulting algorithm a broadened CME, or BCME, approach. 

	With this background in mind, we have two goals for the present article.  
	First, we would like to investigate the {\em consequences and signatures of nonadiabatic effects} for a diatomic molecule scattering off of a metal surface. Experimental signatures of nonadiabatic dynamics have been suggested by Wodtke {\em et al}, and we would like to see how many of these signatures can be studied theoretically.
	To isolate these dynamical effects, we will work with a 2D model that will allow a thorough analysis.
	Second, to guide our understanding of the relevant process, we would also like {\em to compare and contrast three different nonadiabatic dynamics approaches:}
	$(i)$ EF
	$(ii)$ CME
	and $(iii)$ BCME. 
	Because we do not have an exact propagator, it is essential that we analyze multiple approaches. Naturally, since the EF and CME algorithms are based on perturbation theory, these algorithms must be accurate within their own, respective, parameter regimes.
	However, in a non-perturbative regime, it will be crucial to have different approaches so that we can make the best guess for the correct answer.
	In the course of our results, we will point out several surprising features that arise from these different methods. At present, our hypothesis is that, of the three methods above, BCME dynamics are the most reliable.

	This paper is organized as follows: 
	in \refsec{sec: theory} we review all three dynamics schemes discussed above; 
	in \refsec{sec: simulation_details} we define our 2D model Hamiltonian and provide details  of the simulation;
	in \refsec{sec: results}, we show simulation results for different sets of  parameters; 
	in \refsec{sec: discussion}, we discuss the results and highlight why sometimes EF can yield vibrational relaxation rates that are too small while at other times EF can yield vibrational relaxation rates that are too large; 
	in \refsec{sec: conclusion} we conclude with a few suggestions for future work.

	\subsection{Notation} 

	The notations used in this paper are as follows: 
	bold characters (e.g. $\bm{r}$) are vectors,
	bold characters with a left-right arrow (e.g. $\overleftrightarrow{\bm{\Lambda}}$) are tensors, 
	plain characters (e.g. H) are either scalars or operators;
	for indices, we use Greek letters ($\alpha, \beta$, etc) for nuclei and Roman letters ($i,j$, etc) for electrons;  
	$\bm{r}$ and $\bm{p}$ always represent position and momentum vectors for molecules, respectively.

\section{Theory}\label{sec: theory}

	Henceforward, we will consider an idealized molecule (or impurity) on a metal surface using the Anderson-Holstein(AH) model in a nuclear $D$-dimensional space
		\begin{equation}\label{eq: AH_model}
		\begin{aligned}
			H &= \sum_{\alpha=1} ^D \dfrac{p_\alpha^2}{2m_\alpha} + U_0(\bm{r}) + h(\bm{r})d^\dagger d \\
			&\ \ \ \ + \sum_k\epsilon_kc_k^\dagger c_k + \sum_k V_k(\bm{r})(d^\dagger c_k + c_k^\dagger d)
		\end{aligned}
		\end{equation}
	Here and below, the Fermi level of the metal is always chosen to be zero.
	$d$ and $d^\dagger$ are annihilation and creation operators for the impurity site, 
	$c_k$ and $c_k^\dagger$ are annihilation and creation operators for the $k^{th}$ orbital in the electronic bath.
	When $\avg{d^\dagger d} = 1$, the impurity is occupied and we will speak of the molecule as being an anion; 
	when $\avg{d^\dagger d} = 0$, the impurity is unoccupied and we will speak of the molecule as being neutral.
	We define $U_i(\bm{r})$ as the potential energy surface for electronic state $\ket{i}$ at position $\bm{r}$, so that $U_1(\bm{r}) \equiv h(\bm{r}) + U_0(\bm{r})$ is the potential energy surface for the anion. {\em Vice versa}, $h(\bm{r}) \equiv U_1(\bm{r}) - U_0(\bm{r})$ is the energy gap between the anion state and the neutral state.
	$V_k(\bm{r})$ denotes the coupling between the $k^{th}$ bath orbital and the impurity site. 
	$\epsilon_k$ is the energy of the $k^{th}$ bath orbital. 
	Whenever possible, we apply the wide band approximation (WBA) and assume that the self energy of the impurity has only an imaginary part which does not depend on $k$ or $\epsilon$:
		\begin{equation}
		\begin{aligned}
			\Sigma(\epsilon, \bm{r}) &\equiv \sum_k \dfrac{V_k^2(\bm{r})}{\epsilon - \epsilon_k + i\eta} \\
			&\approx -i\pi V^2(\bm{r})\rho(\epsilon) = -i\dfrac{\Gamma(\bm{r})}{2}
		\end{aligned}
		\end{equation}
	Here, $\rho(\epsilon)$ is the density of states in the metallic bath.
	
	\subsection{Electronic Friction} \label{sec: theory_EF}

	In this paper, we will study the dynamics of the AH model using several approaches, especially electronic friction (EF)\cite{HeadGordon1995electronicfriction,Brandbyge1995elefric}. According to this model, all nonadiabatic effects are wrapped up into stochastic Langevin dynamics on a potential of mean force.
	Thus, the equations of motion are:
		\begin{equation}\label{eq: langevin_equation}
		\begin{aligned}
			\partial_t\bm{r} &= \overleftrightarrow{\bm{M}}^{-1}\bm{p} \\
			\partial_t\bm{p} &= \bm{F}(\bm{r}) - \overleftrightarrow{\bm{\Lambda}}(\bm{r})\overleftrightarrow{\bm{M}}^{-1}\bm{p} - \delta \bm{f}(\bm{r}, t)
		\end{aligned}
		\end{equation}
	Here, 
	$\overleftrightarrow{\bm{M}}^{-1} 
		\equiv 
		\begin{bmatrix}
			m_1^{-1} & & \\ 
			& m_2^{-1} & \\
			&	& .. \\
		\end{bmatrix}$ 
	is the inverse mass tensor, 
	$\overleftrightarrow{\bm{\Lambda}}$ is the friction tensor,
	$\delta \bm{f}$ is the random force.
	$\bm{F}$ is the mean force acting on the nuclear degrees of freedom
		\begin{equation} \label{eq: EF_force}
		\begin{aligned}
			\bm{F}(\bm{r}) &= -\nabla U_0(\bm{r}) - \int_{-W}^{W} \bm{K}(\epsilon, \bm{r}) A(\epsilon, \bm{r}) f(\epsilon) d\epsilon
		\end{aligned}
		\end{equation}
	and, for future reference, the potential of mean force is given by 
		\begin{equation}\label{eq: U_PMF}
		\begin{aligned}
			U_{PMF}(\bm{r}) = \zeta(\bm{r_0}) - \int_{\bm{r}_0}^{\bm{r}} \bm{F(\bm{r'})} \cdot d\bm{r'}
		\end{aligned}
		\end{equation}
	where $\zeta(\bm{r}_0)$ is some arbitrary reference potential.
	
	In \refeq{eq: EF_force} the spectral function $A$ and Fermi function $f$ are:
		\begin{equation} \label{eq: A_and_f}
		\begin{aligned}
			A(\epsilon, \bm{r}) &\equiv \dfrac{1}{\pi}\dfrac{\Gamma(\bm{r})/2}{(\epsilon - h(\bm{r}))^2 + (\Gamma(\bm{r})/2)^2} \\
			f(\epsilon) &\equiv \dfrac{1}{e^{\epsilon/kT} + 1}
		\end{aligned}
		\end{equation}

	The kernel $\bm{K}$ in \refeq{eq: EF_force} can be easily computed as\cite{Dou2017nonconstGammaEF}
		\begin{equation} \label{eq: Kernel_vector}
		\begin{aligned}
			\bm{K}(\epsilon, \bm{r}) &\equiv \nabla h(\bm{r}) + (\epsilon - h(\bm{r}))\dfrac{\nabla\Gamma(\bm{r})}{\Gamma(\bm{r})}\\
		\end{aligned}
		\end{equation}

	The reader may well be surprised that the bandwidth $W$ appears in \refeq{eq: EF_force}, given that we would like to take the wide-band limit. In fact, a finite $W$ is required in this case to make sure that the integral in \refeq{eq: EF_force} does not diverge given the form of $\bm{K}$ in \refeq{eq: Kernel_vector} \cite{Dou2016hesterisis}. In practice, we choose $W \gg \Gamma$ (or, to be specific, $W$ is at least 10 times larger than $\Gamma_0$ (see \refeq{eq: gamma_shape})). 

	For a two-state model, with multiple nuclear degrees of freedom, the proper electronic friction tensor is \cite{Dou2017nonconstGammaEF}:
		\begin{equation}\label{eq: EF_fric}
		\begin{aligned}
			\overleftrightarrow{\bm{\Lambda}}(\bm{r}) &= -\pi\hbar\int \left( \bm{K}(\epsilon, \bm{r}) \otimes \bm{K}(\epsilon, \bm{r}) \right) A^2(\epsilon, \bm{r}) f(\epsilon) d\epsilon 
		\end{aligned}
		\end{equation}
	where $\otimes$ denotes an outer product. 

	Finally, $\delta \bm{f}(\bm{r}, t)$ is the random force which is taken to be Markovian and satisfies the fluctuation-dissipation theorem
		\begin{equation}\label{eq: EF_ranforce}
		\begin{aligned}
			\avg{\delta \bm{f}(\bm{r}, t)\otimes\delta \bm{f}(\bm{r}, t')} &= 2kT\overleftrightarrow{\bm{\Lambda}}(\bm{r})\delta(t - t')
		\end{aligned}
		\end{equation}
	Note that \refeq{eq: EF_fric} reduces to the Head-Gordon Tully (HGT) friction model for electronic friction at zero temperature\cite{Dou2017nonconstGammaEF,HeadGordon1995electronicfriction} and is also equivalent to Tully's recent extrapolation for friction at finite temperature\cite{Maurer2016abinit_friction_tensor}.

	From the expressions above, one immediately finds a troubling attribute of electronic friction tensors. For some Hamiltonians, it is possible to encounter geometries where $\Gamma(\bm{r}) \rightarrow 0$ but $\nabla h(\bm{r}) \neq 0$.
 In such a case, the corresponding matrix elements in $\overleftrightarrow{\bm{\Lambda}}(\bm{r})$  (\refeq{eq: EF_fric}) will diverge to infinity
because $\int A^2(\epsilon, \bm{r})d\epsilon \propto \frac{1}{\Gamma}$ as $\Gamma \rightarrow 0$ (see \refeq{eq: A_and_f}). 
	To avoid such a numerical instability, below we will choose a small artificial parameter $\Gamma_{cutoff}$ for our simulations, such that, for $\Gamma(\bm{r}) < \Gamma_{cutoff}$ (corresponding to very small molecule-metal coupling), we will ignore any effect from the electronic bath and set the friction and random force to $0$.
		\begin{equation}\label{eq: Gamma_cutoff}
		\begin{aligned}
			\overleftrightarrow{\bm{\Lambda}}(\bm{r}) &= \text{\refeq{eq: EF_fric}}, \ \ \ \Gamma(\bm{r}) \ge \Gamma_{cutoff} \\
			&= \overleftrightarrow{\bm{0}}, \ \ \ \Gamma(\bm{r}) < \Gamma_{cutoff} \\
		\end{aligned}
		\end{equation}
	We must always check whether or not our final results depend on $\Gamma_{cutoff}$.
	Unless stated otherwise, all data presented below is independent  of $\Gamma_{cutoff}$.

	\subsection{Classical Master Equations} \label{sec: theory_CME}
	Apart from electronic friction, classical master equations (CME) represent an entirely different approach for modeling nonadiabatic dynamics at metal surfaces. The CME approach \cite{Elste2008CME_original, Dou2015CME} treats electronic states explicitly and proposes stochastic trajectories. More specifically, nuclear trajectories are propagated either along $U_0$ or $U_1$ and, for each time step, the particle may hop from one surface to the other. The probability to hop is decided by the hybridization function $\Gamma$. 
	This scheme is summed up by the following equations of motion for the  probability densities:
		\begin{equation}\label{eq: CME}
		\begin{aligned}
			\partial_tP_0(\bm{r}, \bm{p}, t) &= 
			- \overleftrightarrow{\bm{M}}^{-1}\bm{p} \cdot \nabla P_0(\bm{r},\bm{p},t) 
			+ \nabla U_0 \cdot \nabla_{\bm{p}} P_0(\bm{r}, \bm{p}, t) \\
			&\ \ \ \ - \dfrac{\Gamma(\bm{r})}{\hbar}f(h(\bm{r}))P_0(\bm{r}, \bm{p}, t) \\
			&\ \ \ \ + \dfrac{\Gamma(\bm{r})}{\hbar}(1 - f(h(\bm{r})))P_1(\bm{r}, \bm{p}, t) \\
			\partial_tP_1(\bm{r}, \bm{p}, t) &= 
			- \overleftrightarrow{\bm{M}}^{-1} \bm{p} \cdot \nabla P_1(\bm{r},\bm{p},t) 
			+ \nabla U_1 \cdot \nabla_{\bm{p}} P_1(\bm{r}, \bm{p}, t) \\
			&\ \ \ \  + \dfrac{\Gamma(\bm{r})}{\hbar}f(h(\bm{r}))P_0(\bm{r}, \bm{p}, t) \\
			&\ \ \ \ - \dfrac{\Gamma(\bm{r})}{\hbar}(1 - f(h(\bm{r})))P_1(\bm{r}, \bm{p}, t)
		\end{aligned}
		\end{equation}
	Here $P_i(\bm{r}, \bm{p}, t)$ denotes the probability density to find a particle at phase point $(\bm{r}, \bm{p})$ in electronic state $\ket{i}$ at time $t$. 

	The CME in \refeq{eq: CME} can be derived by assuming $(i)$ a high temperature such that classical nuclear motion suffices, and $(ii)$ a small hybridization function  $\Gamma < kT$. Note that, for large enough $\Gamma$ (but not too large) with many hops back and forth between surfaces, CME dynamics become equivalent to EF dynamics\cite{Dou2015CME}. 

	\subsection{Broadened Classical Master Equations} \label{sec: theory_BCME}

	Finally, the last dynamics approach studied here will be an extrapolation of EF and CME dynamics, denoted a broadened CME (BCME)\cite{Dou2016bCME, Dou2016manybodyEF} approach.	According to BCME dynamics, one  modifies the potential surfaces of the different diabatic states so as to include broadening effects. In practice, this modification implies that the diabatic surfaces now depend on both $\Gamma(\bm{r})$ and temperature. 
	The BCME equations of motion are:
		\begin{equation}\label{eq: BCME}
		\begin{aligned}
			\partial_tP_0(\bm{r}, \bm{p}, t) &= 
			-\overleftrightarrow{\bm{M}}^{-1}\bm{p} \cdot \nabla P_0(\bm{r},\bm{p},t) \\ 
			&\ \ \ \ + \left(\nabla U_0 - \Delta\bm{F}^{BCME}(\bm{r}) \right) \cdot \nabla_{\bm{p}} P_0(\bm{r}, \bm{p}, t) \\
			&\ \ \ \ - \dfrac{\Gamma(\bm{r})}{\hbar}f(h(\bm{r}))P_0(\bm{r}, \bm{p}, t) \\
			&\ \ \ \ + \dfrac{\Gamma(\bm{r})}{\hbar}(1 - f(h(\bm{r})))P_1(\bm{r}, \bm{p}, t) \\
			\partial_tP_1(\bm{r}, \bm{p}, t) &= 
			-\overleftrightarrow{\bm{M}}^{-1}\bm{p} \cdot \nabla P_1(\bm{r},\bm{p},t) \\
			&\ \ \ \ + \left(\nabla U_1 - \Delta\bm{F}^{BCME}(\bm{r}) \right) \cdot \nabla_{\bm{p}} P_1(\bm{r}, \bm{p}, t) \\
			&\ \ \ \ + \dfrac{\Gamma(\bm{r})}{\hbar}f(h(\bm{r}))P_0(\bm{r}, \bm{p}, t) \\
			&\ \ \ \ - \dfrac{\Gamma(\bm{r})}{\hbar}(1 - f(h(\bm{r})))P_1(\bm{r}, \bm{p}, t) \\
		\end{aligned}
		\end{equation} 
	Here, the diabatic forces have been modified by the following correction
		\begin{equation}\label{eq: delta_F_BCME}
		\begin{aligned}
			\Delta\bm{F}^{BCME}(\bm{r}) &=  f (h(\bm{r}))\nabla h(\bm{r}) \\
				&\ \ \ \ - \int_{-W}^{W} \bm{K}(\epsilon, \bm{r}) A(\epsilon, \bm{r}) f(\epsilon) d\epsilon \\
				&= -\nabla h(\bm{r}) (n(\bm{r}) - f(h(\bm{r}))) \\
				&\ \ \ \ - \int_{-W}^{W} \left((\epsilon - h(\bm{r}))\dfrac{\nabla\Gamma(\bm{r})}{\Gamma(\bm{r})}\right) A(\epsilon, \bm{r}) f(\epsilon) d\epsilon \\
			n(\bm{r}) &\equiv \int_{-W}^{W} A(\epsilon, \bm{r}) f(\epsilon) d\epsilon \\
		\end{aligned}
		\end{equation}
	For future reference the broadened diabatic potentials of mean force are:
		\begin{equation}\label{eq: U_BCME}
		\begin{aligned}
			U_i^b(\bm{r}) \equiv U_i(\bm{r}) - \int_{\bm{r}_0}^{\bm{r}} \Delta\bm{F}^{BCME}(\bm{r'}) \cdot d\bm{r'} + \zeta(\bm{r_0}),  \ \ i = 0,1
		\end{aligned}
		\end{equation}
	In \refeq{eq: delta_F_BCME}, $f$ is the Fermi function and $f(h(\bm{r}))$ represents the unbroadened, equilibrium population for the impurity site at position $\bm{r}$. 
	By contrast, $n$ represents the correctly broadened equilibrium population of  for the impurity site at position $\bm{r}$. 
	Thus, $n-f$ indeed represents a broadening correction. We note that, for large enough $\Gamma$, the total probability density for BCME dynamics evolves on the same potential of mean force as EF dynamics (in \refeq{eq: U_PMF})\cite{Dou2016bCME}. Below, in section \refsec{sec: results_dynamics}, we will plot and compare the unbroadened ($U_i$) and broadened ($U_i^b$) diabats.

\section{Simulation Details} \label{sec: simulation_details}

	To study the methods above, we will simulate vibrational relaxation for a model two-dimensional (2D) system. 
	Our 2D system has been roughly designed to mimic a scattering event whereby a diatomic molecule impinges on a metal surface. The first dimension $x$ corresponds to the vibrational DoF of the molecule, and the second dimension $z$ is the molecular center-of-mass position. 
	The energy surfaces we use are
		\begin{equation}\label{eq: surfaces_shape}
		\begin{aligned}
			U_0(\bm{r}) &= R_0(x) + S_0(z) \\
			U_1(\bm{r}) &= R_1(x) + S_1(z) \\
			R_0(x) &= \dfrac{1}{2}m_x\omega^2(x - x_0)^2 \\
			R_1(x) &= \dfrac{1}{2}m_x\omega^2(x - x_1)^2 \\
			S_0(z) &= A_0(e^{2C_0(z - z_0)} - 2e^{C_0(z - z_0)})\\
			S_1(z) &= \dfrac{1}{4z} + \dfrac{1}{(z - C_1)^6} + B_1
		\end{aligned}
		\end{equation}
	Here $m_x$ is the reduced mass for vibrational motion.
	The energy surfaces along the $x$ direction are harmonic wells, where the eigenfrequency $\omega$ is chosen to be the same for $\ket{0}$ and $\ket{1}$. 
	The energy surfaces in the $z$ direction  ($S_0(z), S_1(z)$) resemble the electron mediated model for NO proposed by  Newns\cite{Newns1986NOsurfacemodel}. 
	The second term for motion in the expression for $S_1(z)$) does not appear in Newns model, but has been added to ensure that the impinging NO particles scatter back (rather than penetrate the metal). 
	The metal surface is effectively located around $z = 0$.

	For the hybridization function $\Gamma(\bm{r})$, we choose:
		\begin{equation} \label{eq: gamma_shape}
		\begin{aligned}
			\Gamma(\bm{r}) &= \Gamma_0 Q(x) T(z) \\
			Q(x) &= 1 + e^{-K_gx^2} \\
			T(z) &= \dfrac{2}{1 + e^{c_g(z - z_g)}}
		\end{aligned}
		\end{equation}
	In the $x$ direction, $\Gamma(\bm{r})$ has a maximum near the equilibrium position of the $\ket{0}$ state (i.e. $x_0$); 
	in the $z$ direction, $\Gamma(\bm{r})$ decreases exponentially as the distance between particle and surface increases, and $\Gamma(\bm{r})$ goes to 0 as $z  \rightarrow -\infty$.  

	Almost all of parameters listed above are defined in \reftable{table: parameters_in_vib} (except for the hybridization function $\Gamma_0$ and the displacement $x_1$). These parameters were chosen to (very roughly) simulate the scattering of NO from an Au surface. 
	Note that the temperature here is relatively low, and should not satisfy the ``high temperature'' prerequisite for CME dynamics. That been said, the experiments start in a hot vibrational state $n_{vib} = 15$ (which makes the classical vibrational energy $E_{vib} \gg \hbar\omega$), such that classical dynamics may well still be valid. Furthermore, in this paper we will also study the dynamics with BCME to include broadening.  (In a future publication, we will consider these dynamics with a broadened version of the QME (to include broadening plus nuclear quantum effects.)
	For now, our major concern is how will the dynamics depend on different values of $x_1$ and $\Gamma_0$ (as well as in the incoming momentum in the $z$-direction, $\avg{p_0}$).
	
	\begin{table}[ht]
		\centering 
		\caption{Parameters used in the simulation} 
		\begin{tabular}{c c c c } 
			\hline
			Parameter & Value(a.u.) & \ \ \ \  &Comment\\ 
			\hline 
			$m$ & 55000 &  & mass of particle\\ 
			$m_x$ & 14000 & & reduced mass \\
			$kT$ & 0.001 & & temperature \\
			$\omega$ & 0.008 & & harmonic frequency\\
			$x_0$ & 0 & & parameter in \refeq{eq: surfaces_shape} \\
			$A_0$ & 0.011 & & parameter in \refeq{eq: surfaces_shape}\\
			$C_0$ & 0.64 & & parameter in \refeq{eq: surfaces_shape}\\
			$z_0$ & -3.5 & & parameter in \refeq{eq: surfaces_shape}\\
			$B_1$ & 0.2 & & parameter in \refeq{eq: surfaces_shape}\\
			$C_1$ & 0.67 & & parameter in \refeq{eq: surfaces_shape}\\
			$K_g$ & 4 & & parameter in \refeq{eq: gamma_shape}\\
			$c_g$ & 0.64 & & parameter in \refeq{eq: gamma_shape}\\
			$z_g$ & 0 & & parameter in \refeq{eq: gamma_shape}\\
			$W$ & 1.5	& & bandwidth in \refeq{eq: EF_force}\\
			\hline 
		\end{tabular}
		\label{table: parameters_in_vib} 
	\end{table}
	
	\begin{figure}[ht]
		\centering
		\includegraphics[width=5in]{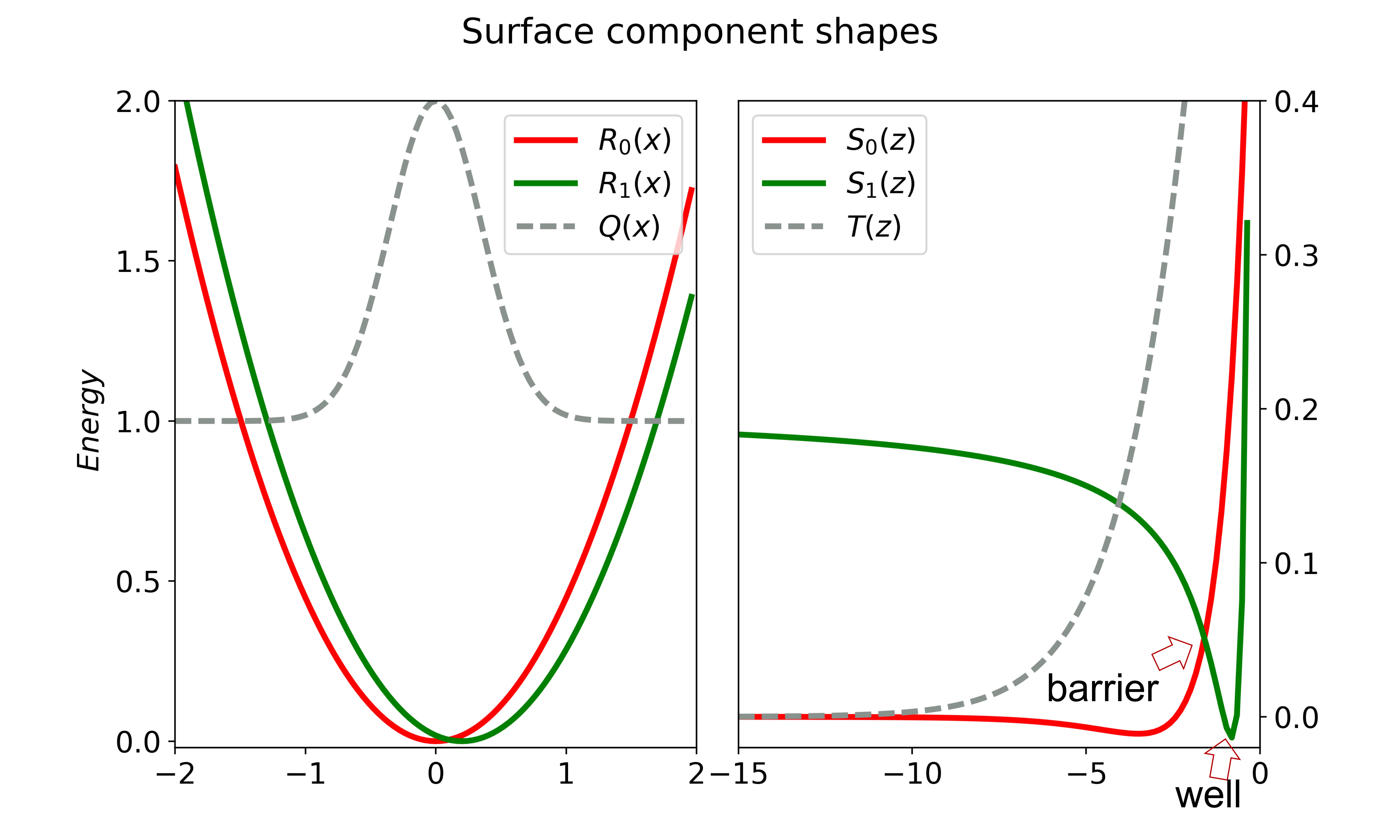}
		\caption{ A plot of the 2D model surfaces used in our simulation. The energy surfaces in the $x$ direction (left) are both harmonic wells. The equilibrium positions for the neutral state and the anion state are $x_0=0$ and $x_1=0.2$  (see \refeq{eq: surfaces_shape}). In the $z$ direction (right),  there is an energy barrier at $z = -1.8$ and an energy well at $z = -0.9$ (see arrows in the plot), which can potentially trap incoming particles. }
		\label{fig: surf_for_vib_anal}
	\end{figure}
	
	\begin{figure}[ht]
		\centering
		\includegraphics[width=4in]{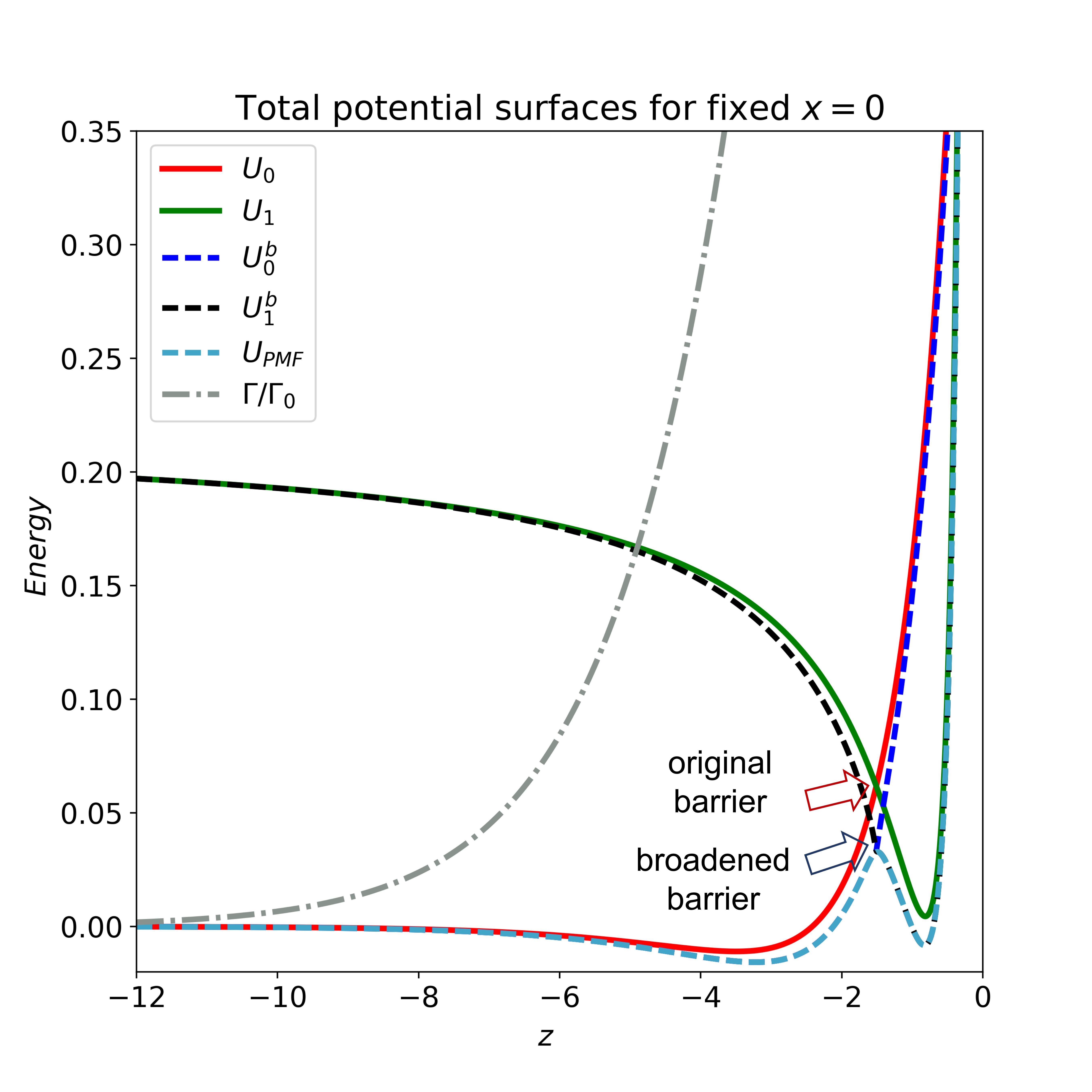}
		\caption{ The CME and BCME surfaces in the $z$-direction. Here $x = 0.0$,  $\Gamma_0 = 0.03$. $U_i$ and $U_i^b$ are the CME and BCME surfaces for $\ket{i}$, respectively. $U_{PMF}$ is calculated according to \refeq{eq: U_PMF}. For a particle incoming from $z\rightarrow -\infty$, on the lower surfaces $U_0$, there is an energetic barrier to reach the crossing point. When broadening is taken into account (e.g. through the BCME), this barrier is lowered. The arrows show the crossing point with or without broadening. }
		\label{fig: broadened_surf}
	\end{figure}
	
	In \reffig{fig: surf_for_vib_anal}, we plot the individual components making up the diabatic potential energy surfaces in \refeq{eq: surfaces_shape}.
	In \reffig{fig: broadened_surf}, we plot the total the potential surfaces in the $z$-direction (for one fixed $x$), and we show the effects of broadening. 
	We also plot the hybridization function $\Gamma(\bm{r})$. 

	For each calculation reported below, we have run 5000 trajectories. 
	To roughly simulate the Wodtke experiments\cite{Huang2000NOAuexp},  each trajectory was initialized to the $15^{th}$ vibrational state. 
	In the $x$ direction, trajectories were initialized with a microcanonical ensemble: we weighted all $(x, p_x)$ satisfying $E_x = E_{vib}^0 = 15.5 \hbar\omega$ equally.
	In the $z$ direction, trajectories were initialized at position $z = -15$ and the momentum $p_z$ was chosen from a Gaussian distribution with $\avg{p_0}$ and $\sigma = \sqrt{mkT_{0}}, kT_{0} = 0.5kT$. 
	We used a time step $dt = 0.25$ a.u. and propagated trajectories for $2\times10^{6}$ steps.
	Dynamics were carried out with the velocity-verlet propagator. 
	Unless stated otherwize, trapped particles were not considered and we analyzed exclusively reflected particles (which were collected at $z = -20$).  

\section{Results}\label{sec: results}

	We will now report our results, focusing mostly on the overall amount of predicted vibrational relaxation.

	\subsection{Dynamics} \label{sec: results_dynamics}

\begin{figure}[ht]
		\centering
		\includegraphics[width=3in]{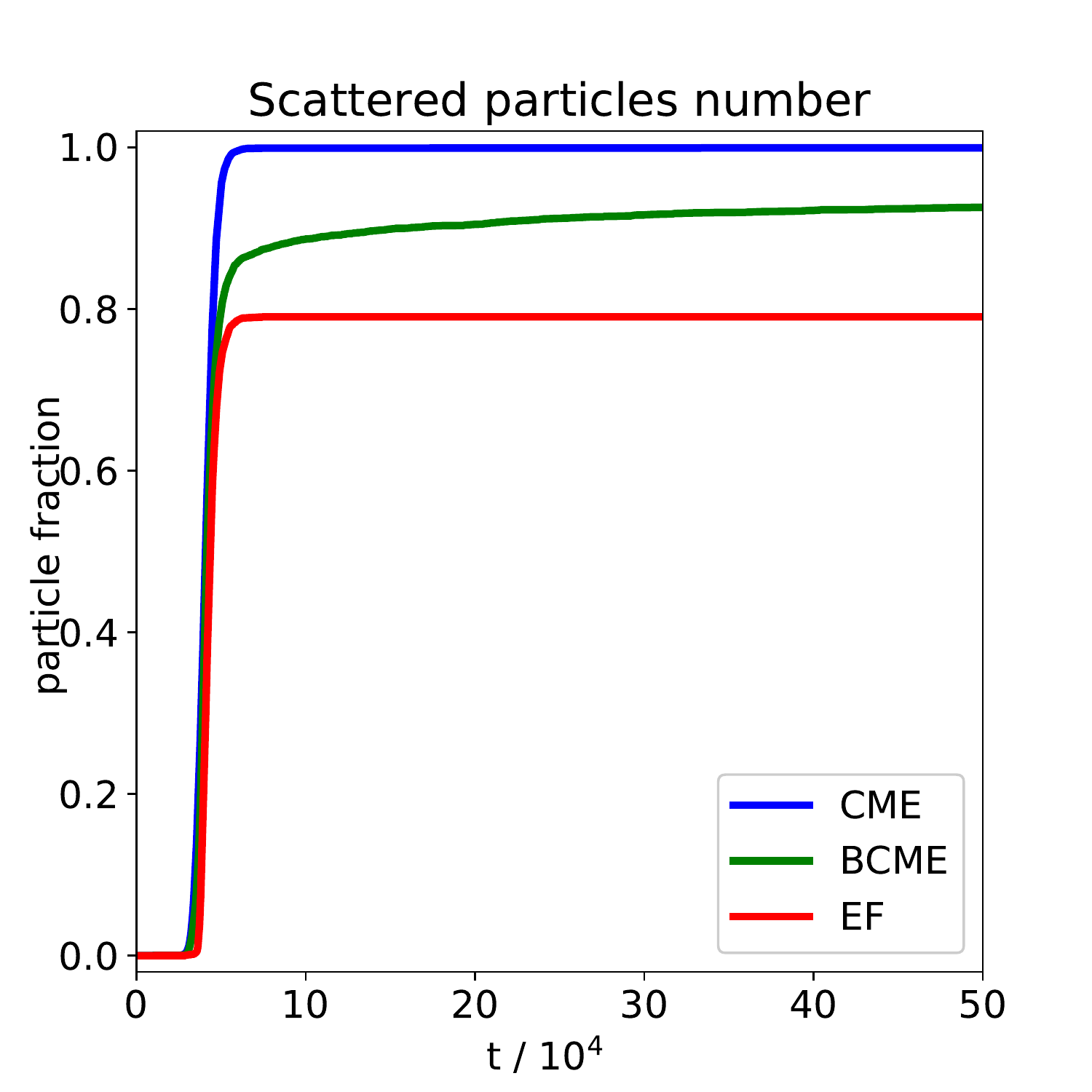}
		\caption{ Fraction of scattered particles as a function of $t$. Here $x_1 = 0.2$, $\Gamma_0 = 0.03$,$\avg{p_0} = 40$, $\Gamma_{cutoff} = 0.075\hbar\omega$. Note that the CME dynamics results in no trapping, the BCME dynamics result in a modest amount of trapping, and the EF dynamics result in the most trapping (relatively). We have checked that these reported fractions are unchanged for a long time after t = $49\times 10^4$, such that the percentage of trapped trajectories are very meaningful plateau values. }
		\label{fig: Nout_dynamics}
	\end{figure}

	In \reffig{fig: Nout_dynamics}, we plot the number of particles collected at $z = -20$ as a function of time. 
	In this case, for CME dynamics, we find very few particles trapped near the surface.
	For BCME dynamics, $ \sim 10\%$ of the particles are trapped, 
	and for EF dynamics, more than 20\% of the particles trap in the well near $z = -0.9$. 
	In general, we find that this trend holds for most calculations below with different parameters: the EF results usually result in far more trapping that CME or BCME dynamics. (see \reftable{table: scattered_particle_fraction})

	\begin{table}[ht]
		\centering 
		\caption{Scattered Particle Statistics} 
		\begin{tabular}{p{1cm} p{1cm} p{1cm} p{1.5cm} p{1.5cm} p{1.5cm} p{1.5cm} p{1.5cm} p{1.5cm}} 
			\hline
			\multicolumn{3}{c}{parameters}  & \multicolumn{3}{c}{percentage trapped} & \multicolumn{3}{c}{percentage reflected on state $\ket{1}$} \\ 
			\hline 
			$x_1$ & $\Gamma_0$ & $\avg{p_0}$  &  CME  & BCME & EF & CME  && BCME \\
			\hline 
			$0.2$ & $0.01$ & $20$ & $0.02\%$  & $0.04\%$  & $0.02\%$   & $0.04\%$  && $0.08\%$ \\
			$0.2$ & $0.03$ & $20$ & $0.02\%$  & $1.36\%$  & $0.72\%$   & $0.14\%$  && $0.08\%$ \\
			$0.2$ & $0.05$ & $20$ & $0.02\%$  & $10.24\%$ & $17.44\%$  & $0.06\%$  && $0.11\%$ \\
			$0.2$ & $0.08$ & $20$ & $0.04\%$  & $37.06\%$ & $55.34\%$  & $0.08\%$  && $0.03\%$ \\
			$0.2$ & $0.03$ & $40$ & $0.08\%$  & $7.44\%$  & $20.98\%$  & $0.06\%$  && $0.09\%$ \\
			$0.2$ & $0.03$ & $60$ & $3.60\%$  & $6.26\%$  & $18.50\%$  & $0.10\%$  && $0.09\%$ \\
			$0.2$ & $0.03$ & $80$ & $1.22\%$  & $0.78\%$  & $0.32\%$   & $0.06\%$  && $0.10\%$ \\			
			$0.4$ & $0.01$ & $20$ & $7.98\%$  & $8.52\%$  & $2.06\%$   & $0.04\%$  && $0.07\%$ \\
			$0.4$ & $0.03$ & $20$ & $3.26\%$  & $4.42\%$  & $2.24\%$   & $0.10\%$  && $0.08\%$ \\
			$0.4$ & $0.05$ & $20$ & $1.34\%$  & $3.24\%$  & $31.34\%$  & $0.14\%$  && $0.12\%$ \\
			$0.4$ & $0.08$ & $20$ & $0.34\%$  & $1.98\%$  & $7.60\%$   & $0.08\%$  && $0.10\%$ \\
			$0.4$ & $0.03$ & $40$ & $9.38\%$  & $7.48\%$  & $23.22\%$  & $0.09\%$  && $0.06\%$ \\
			$0.4$ & $0.03$ & $60$ & $9.68\%$  & $4.90\%$  & $8.24\%$   & $0.04\%$  && $0.06\%$ \\
			$0.4$ & $0.03$ & $80$ & $3.74\%$  & $1.88\%$  & $5.48\%$   & $0.04\%$  && $0.04\%$ \\
			\hline 
		\end{tabular}
		\label{table: scattered_particle_fraction} 
	\end{table} 
	
	\subsection{Vibrational Distribution} \label{sec: results_vibrational_distrubution}

	Let us now discuss the vibrational relaxation of the outgoing particles that are scattered backwards (and ignore all trapped particles).   
	With CME or BCME dynamics, because of the large energy penalty to emerge as an anion asymptotically, almost all ($> 99\%$) reflected particles are found to lie on the neutral state $\ket{0}$.
	For this reason, the vibrational state of each particle can be calculated  as follows:
	$(a)$ we compute the kinetic energy in the $x$ direction, $E_{kx} = p_x^2 / 2m_x$,
	$(b)$ we compute the potential energy in the $x$ direction,$E_{px} = U_0(x)$,
	$(c)$ we compute $n_{vib} = (E_{kx} + E_{px}) / \hbar\omega - 0.5$, and round it to an integer.
	This procedure can be applied for all methods above (CME, BCME and EF). 
	Note that, for BCME dynamics, we may safely use $U_0(x)$ (rather than $U_0^b(x)$) because at $z = -20$, $\Gamma \rightarrow 0$, and the $U_0(x)$ and the $U_0^b(x)$ surfaces have negligible differences.
	
	\subsubsection{Relaxation Dependence on $\Gamma_0$}

	In \reffig{fig: vib_anal_g}, we plot the vibrational distributions as a function of different $\Gamma_0$ for fixed incoming z-momentum $\avg{p_0} = 20$.  
	We observe vibrational relaxation for both EF and BCME dynamics, while CME dynamics do not yield any relaxation. 
	From these plots, it is straightforward to see that CME dynamics fail for an obvious reason: nearly all particles are blocked by the energy barrier in the $z$ direction, and they do not have enough energy to reach the surface crossing seam, see \reffig{fig: surf_for_vib_anal}.

	Focusing now on EF and BCME dynamics, as one would expect, we find that relaxation becomes stronger as the molecule-bath coupling increases from zero, reaching a maximum($\avg{n_{vib}} = 7-8$) at $\Gamma_0 = 0.05$.  Note, however, that there is a turnover. As $\Gamma_0$ increases even more, corresponding to the extreme adiabatic limit, vibrational relaxation slows down.
	Such a surprising turnover feature is not found in condensed phase dynamics (with an external friction), where the rate of molecule-metal electron transfer is strictly increasing with the coupling parameter $\Gamma$. See, e.g., Fig 2 in Ref \onlinecite{Ouyang2016ele_transfer_EF_CME_compare} and Fig 10.8 in Ref \onlinecite{Schmickler2010InterfacialElectrochemistry}. We believe this turnover is caused by the fact that, as $\Gamma_0$ increases, the maximum value of the friction tensor decreases, and the dynamics become completely adiabatic, and non-adiabatic effects are minimized. At the same time, though, there is obviously no relaxation for $\Gamma_0 = 0$, and thus there must be a turnover.

	Finally, regarding reliability, EF and BCME relaxation rates agree, especially in the adiabatic limit as $\Gamma_0$ increases. This agreement gives us a large amount of confidence in the quality of our data. Even though we cannot propagate exact dynamics, we have now demonstrably calculated similar observables with two different and orthogonal methods.
	In \reffig{fig: vib_anal_g_largex}, however, we show this agreement is not universal. Here, we plot the same result for the displacement $x_1 = 0.4$ (see \refeq{eq: surfaces_shape}), and now we find that EF and BCME are in far worse agreement. While both methods predict more relaxation than the case of $x_1 = 0.2$, the BCME approach predicts far more relaxation for small $\Gamma_0$ than does EF.  
	In this case, because CME dynamics can be derived with perturbation theory assuming small $\Gamma$, it is easy to argue that CME dynamics (and not EF) dynamics	must be accurate here for $\Gamma_0 = 0.01$. 
	Furthermore, from the fact that BCME dynamics exactly agree with CME dynamics for small $\Gamma_0$  and qualitatively agree with EF dynamics for large $\Gamma_0$, we hypothesize that BCME dynamics should be meaningful over a wide range of parameter space. 
	Our intuition is that EF dynamics will fail for large displacements ($x_1 - x_0$) and small or moderately sized hybridization functions $\Gamma_0$.

	\subsubsection{Relaxation Dependence on incoming momentum}
		 
\begin{figure}
			\centering
			\subfloat[]{\includegraphics[width= 3in]{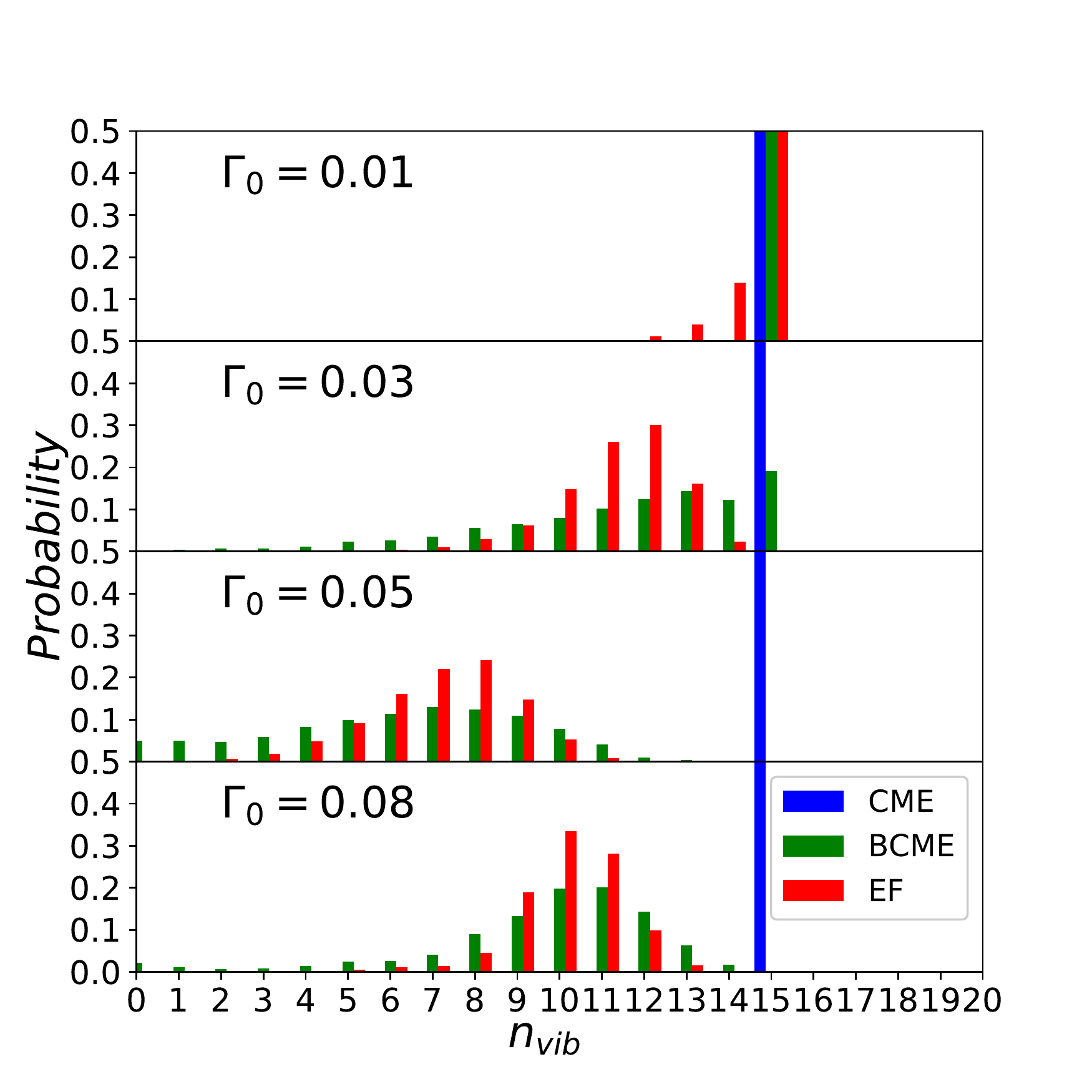}\label{fig: vib_anal_g}}
			\quad
			\subfloat[]{\includegraphics[width= 3in]{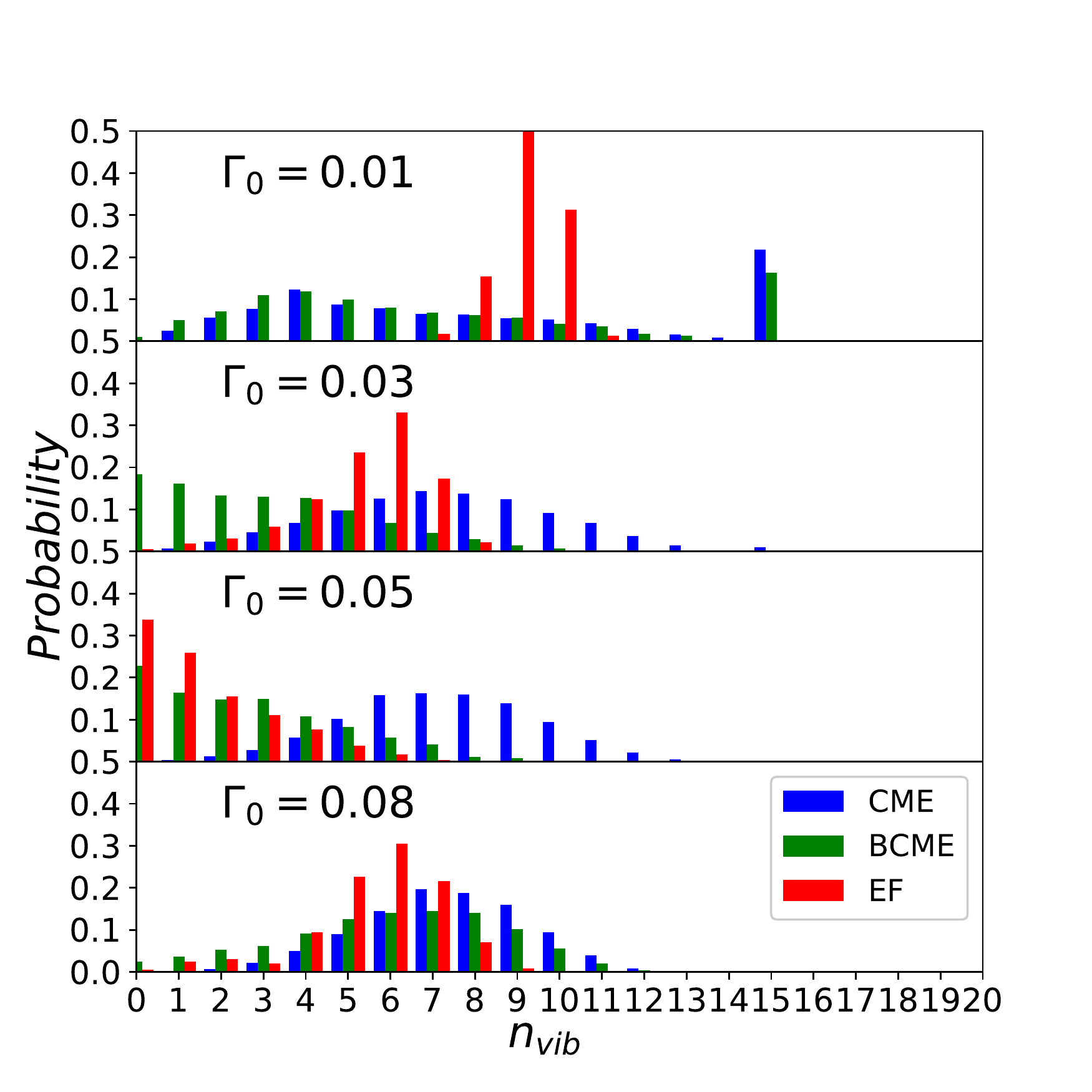}\label{fig: vib_anal_g_largex}}
			\caption{ Vibrational distribution analysis for the scattered trajectories for different metal-molecule coupling. Here $x_1 = 0.2$(left), $0.4$(right) (see \refeq{eq: surfaces_shape}). The incoming z-momentum $\avg{p_0} = 20$, $\Gamma_{cutoff} = 0.075\hbar\omega$, $\Gamma_0 = 0.01, 0.03, 0.05, 0.08$. CME dynamics give no relaxation because without broadening, the trajectories never reach the diabatic crossing point. However, both EF and BCME give relaxation, and the methods agree for large $\Gamma_0$. Note that both of these methods predict a turnover in relaxation: vibrational relaxation is maximized for $\Gamma_0 = 0.05$. Note also that, for $x_1 = 0.4$, BCME gives significantly more relaxation than EF when $\Gamma_0$ is small. }
			\label{vib_anal_G}
		\end{figure}

\begin{figure}
			\centering
			\subfloat[]{\includegraphics[width= 3in]{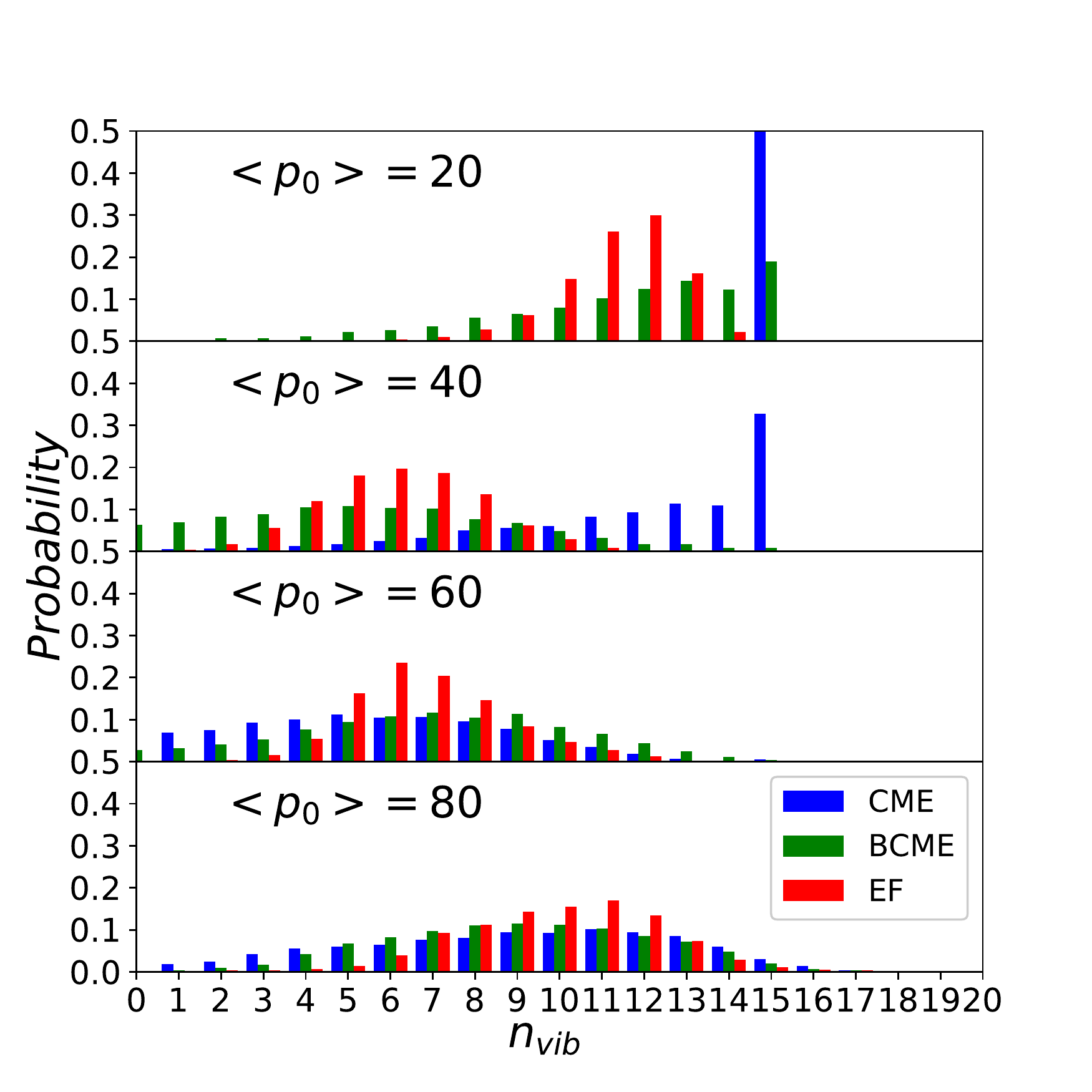}\label{fig: vib_anal_v}}
			\quad
			\subfloat[]{\includegraphics[width= 3in]{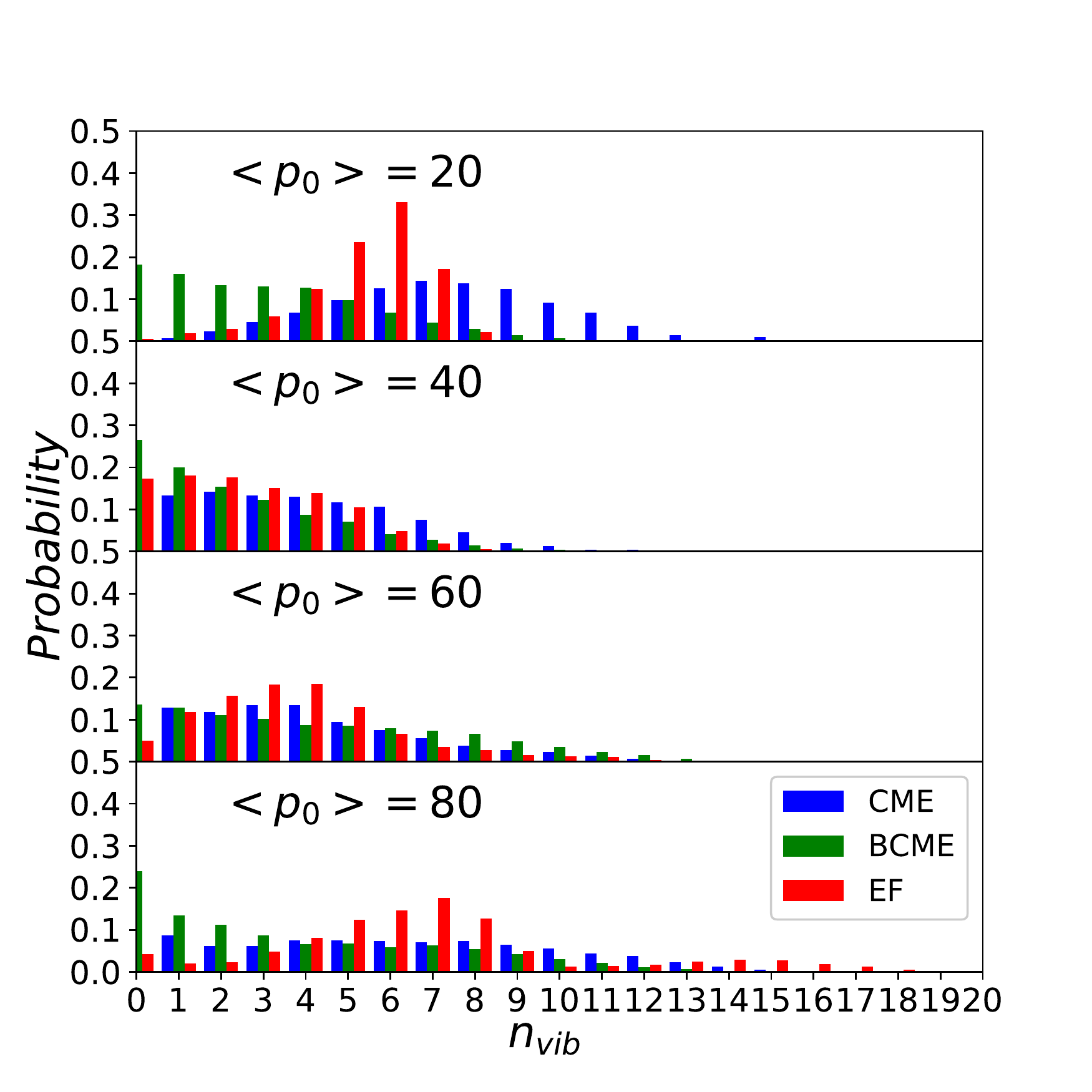}\label{fig: vib_anal_v_largex}}
			\caption{ Vibrational distribution analysis for the scattered trajectories for different incoming momenta. Here $x_0 = 0.2$(left), $0.4$(right), $\Gamma_0 = 0.03$, $\Gamma_{cutoff} = 0.075\hbar\omega$, $\avg{p_0} = 20, 40, 60, 80$. The agreement between EF and BCME increases as $\avg{p_0}$ increases. When $x_1 = 0.4$, more relaxation is predicted for all three schemes. }
			\label{vib_anal_V}
		\end{figure}

	We now study how the incoming momentum affects relaxation.
	In \reffig{fig: vib_anal_v}, we plot vibrational distributions for different $\avg{p_0}$ with a  fixed value of the hybridization ($\Gamma_0 = 0.03$) and $x_1 = 0.2$. Here, for $\avg{p_0} \ge 40$, the CME approach finally gives relaxation (compared against \reffig{fig: vib_anal_g}): there is enough energy to reach the diabatic crossing point. However, the CME does not agree with BCME or EF dynamics for small momenta. Regarding EF and BCME dynamics, we find that the relaxation rates are also in disagreement (though not completely different) for small incoming momenta. 

	For large momenta, however, we note that all dynamic protocols (EF, CME, BCME) roughly agree: apparently, because of the large incoming momenta, there are enough classical crossings such that friction results become meaningful but this kinetic energy is also large enough such that broadening effects on the surface are unimportant. This agreement between CME and EF dynamics has been seen before in 1D problems\cite{Ouyang2016ele_transfer_EF_CME_compare}.

	Lastly, we consider the same dynamics now for the case of a larger displacement, $x_1 = 0.4$. 
	Here, we find again that there is no agreement between any of the methods for small incoming momentum. Because of its ability to interpolate, however, we hypothesize that BCME dynamics are the most accurate. That being said, at larger incoming velocities, the methods do become more similar.  
	Interestingly, though, at very large incoming velocities, all methods become very different again. These features cannot yet be easily explained. 
	In general, we find that BCME dynamics consistently predict more relaxation than electronic friction as well as slightly wider vibrational distributions.
	
	\subsection{Electronic Energy Released and Hopping Energy Histograms} \label{sec: results_enegry_histogram}

\begin{figure}
		\centering
		\includegraphics[width= 5in]{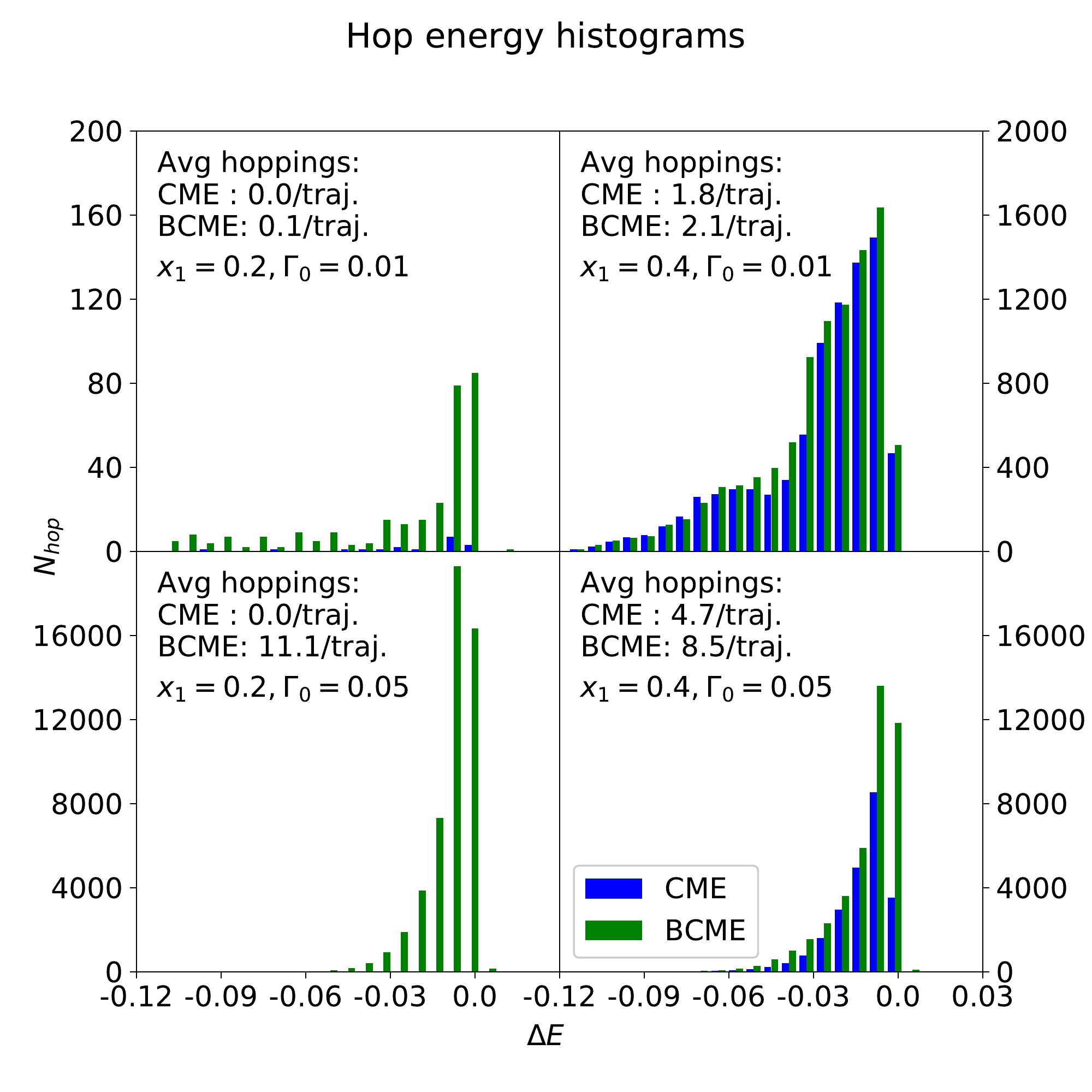}
		\caption{ A histogram of hopping energies. Here $x_1 = 0.2, 0.4$, $\Gamma_0 = 0.01, 0.05$, $\avg{p_0} = 20 $, $\Gamma_{cutoff} = 0.075\hbar\omega$. Positive $\Delta E$ means energy transfers from the metal to a trajectory, while negative $\Delta E$ means energy transfers from a trajectory to the metal. Note that the $y$ axis has different scales for different subfigures. The plot suggests that, for all 4 cases, most hops have small energy changes, but a large energy transfer is not prohibited. Note the sensitivity of the BCME dynamics to $x_1$.}
		\label{fig: Nhop_compare}
	\end{figure}

	Recent experiments have actually measured the distribution of electronic kinetic energies excited in a metal surface as the result of molecular scattering\cite{Larue2011NOAu_kinetic_energy_distribution,Larue2011NOAu_kinetic_energy_distribution2}. With this experimental fact in mind, we plot the energy distribution for hopping according to CME/BCME dynamics for the four different simulations in \reffig{fig: Nhop_compare}. Note that such energy distributions cannot easily be extracted from EF calculations.

	The results in \reffig{fig: Nhop_compare} show that, for most hops, energy transfers from the incoming particle to the metal (i.e. the particle loses energy). Most hops occur near the surface crossing region with small energy gaps ($|\Delta E| < 0.02$). Even so, large energy transfer events are possible within a single hopping event, which does explain the ``multi-quanta relaxation" observed in Wodtke's experiments\cite{White2005NOAuExp_vibrelax,Wodtke2016Rev}. These results will be discussed in the next section.

\section{Discussion}\label{sec: discussion}

	Thus far, we have found that both electronic friction and broadened classical master equation are able to capture many features of vibrational relaxation, and sometimes these two methods even agree. 
	At this point, however there are two key features which must be discussed in more detail.
	
	\subsection{BCME is more sensitive to displacement than EF} 

	From \reffig{vib_anal_G} and \reffig{vib_anal_V}, we observed that, although EF and BCME both yield larger relaxation rates when the displacement $x_1$ is increased from $0.2$ to $0.4$, the BCME approach is obviously more sensitive to this change in parameter -- especially for small $\Gamma_0$ and small $\avg{p_0}$ cases.	This sensitivity is obviously important because, for many molecules, the anionic and neutral potential energy surfaces can be very different. Furthermore, EF should be reliable only when these differences (i.e. electron-phonon couplings) are relatively small. 
	
\begin{figure}
		\centering
		\subfloat[]{\includegraphics[width= 3in]{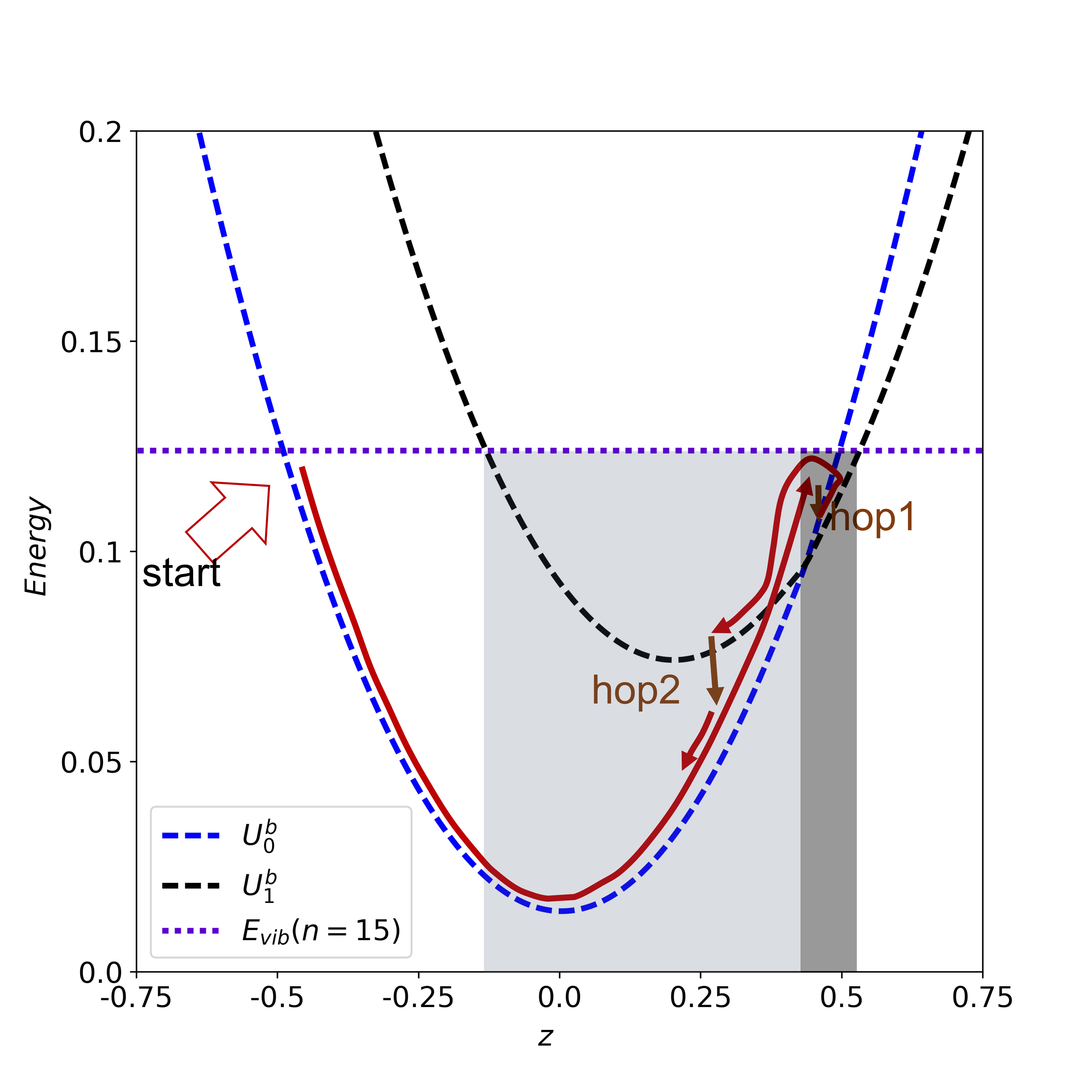}\label{fig: hop_scheme}}
		\quad
		\subfloat[]{\includegraphics[width = 3in]{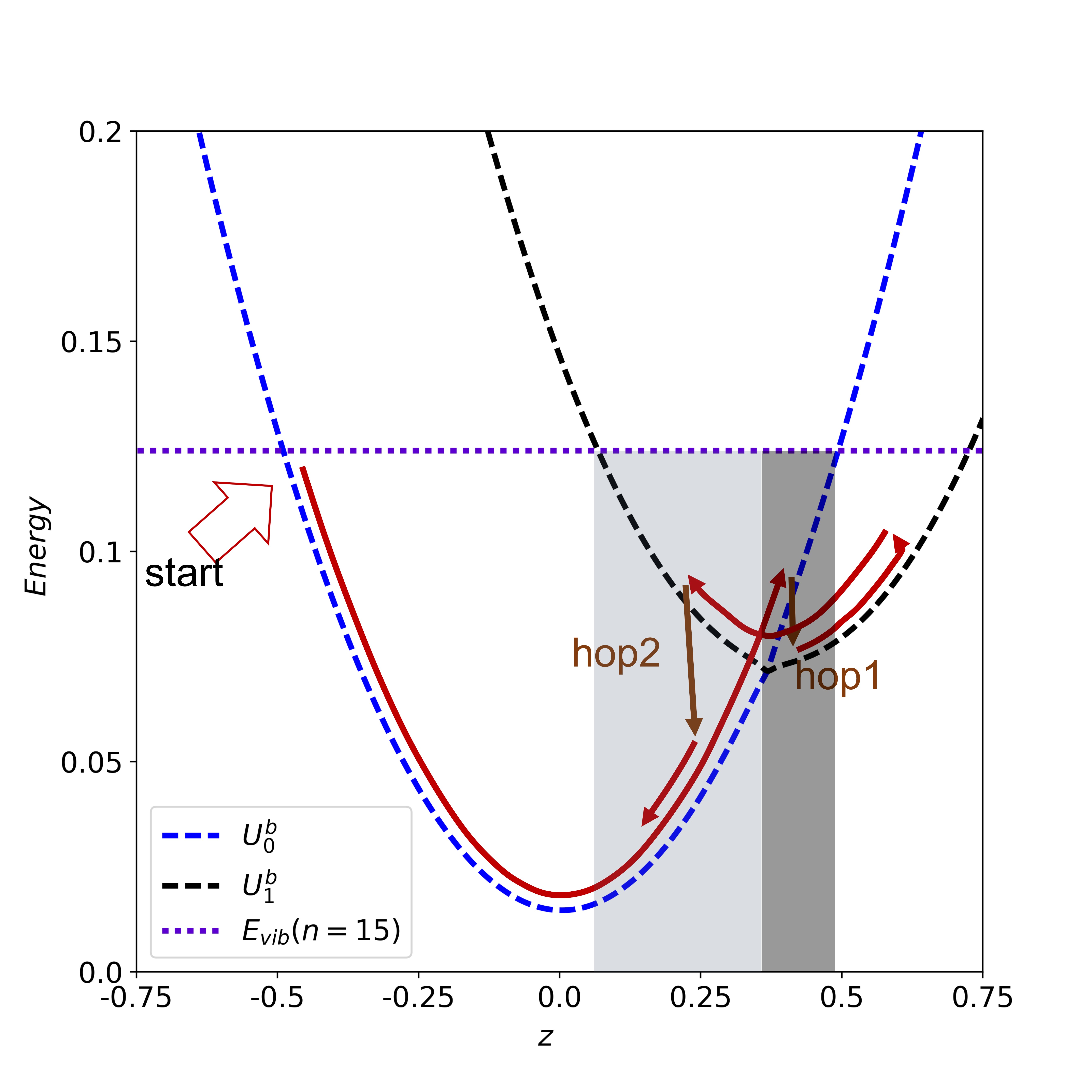}\label{fig: hop_scheme_largex}}
		\caption{ BCME surfaces in the $x$ direction for $x_1 = 0.2$(left), $0.4$(right), with a fixed $z=-2.0$. Here $\Gamma_0 = 0.01$. The dotted line is the vibrational energy for particles at $n_{vib} = 15$, and the shaded areas are the active regions for hopping events $\ket{0}\rightarrow\ket{1}$ (darker) and $\ket{1}\rightarrow\ket{0}$ (lighter), assuming only downward hops(suggested by \reffig{fig: Nhop_compare}). Because particles move more slowly in the $z$-direction than in the $x$-direction when $n_{vib} = 15$, this cartoon representation (with fixed $z$ position) gives a reasonable explanation for why vibrational relaxation is faster with $x_1 = 0.4$ (as opposed to $x_1 = 0.2$). Obviously, $hop1$ can be triggered more easily when $x_1 = 0.4$, and $hop1$ also releases more vibrational energy to the metal for larger $x_1$. }
		\label{hop_schemes}
	\end{figure}

	To explain the sensitivity of BCME dynamics, a figure will be very useful (\reffig{hop_schemes}). Here, we observe that, as the displacement $x_1$ becomes larger, the surface crossing point as a function of the $x$ coordinate drops in energy. As a result, if trajectories move along diabats, trajectories with  a given $n_{vib}$ will spend more time in regions of large hopping probability. Furthermore, in these very regions, there is the chance to lose a larger amount of energy in one hop (see also \reffig{fig: Nhop_compare}).

	Now, EF dynamics also predict stronger relaxation for large displacements -- after all, the EF friction tensor is proportional to $\nabla h$ (see \refeq{eq: Kernel_vector} and \refeq{eq: EF_fric}). However, EF dynamics are not as sensitive to the displacement as are BCME dynamics because EF dynamics move along the adiabatic surface (rather than the diabatic surface) and a dramatic, sudden energy loss is impossible. Indeed, Wodtke and Tully {\em et al} have argued that EF dynamics cannot produce multi-quanta relaxation because, by damping the nuclear motion, nuclear velocities change continuously time, and thus any quantum mechanical extension of electronic friction must predict step-by-step dissipation of vibrational quanta\cite{Wodtke2016Rev,Shenvi2006NOAuTheory_vibrelax_model}.

	\subsection{EF can be sensitive to $\Gamma_{cutoff}$}

\begin{figure}
		\centering
		\includegraphics[width=5in]{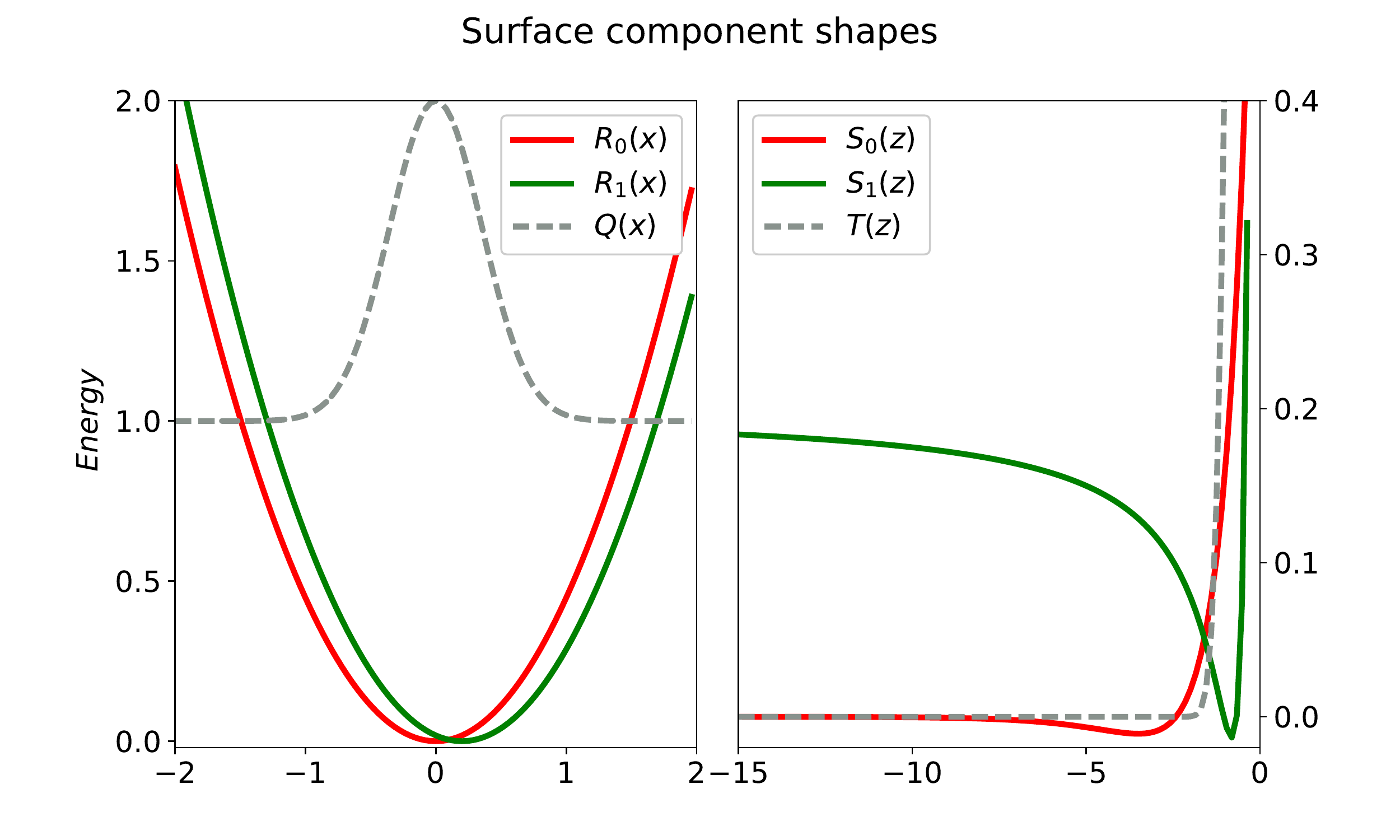}
		\caption{ Modified surfaces used in $\Gamma_{cutoff}$ analysis (\reffig{vib_anal_CUT}). $x_1 = 0.2$. The only difference between this figure and \reffig{fig: surf_for_vib_anal} is the shape of $T(z)$, which is relatively sharper here. Here  $\Gamma(\bm{r})$ can be very small (but still $> \Gamma_{cutoff}$) even in the surface crossing region. }
		\label{fig: surf_for_cutoff_anal}
	\end{figure}

	The very last feature that must be discussed is the  artificial parameter, $\Gamma_{cutoff}$, which we have included above (in \refeq{eq: Gamma_cutoff}) so as to determine when to apply or not apply the frictional damping and random force. The parameter $\Gamma_{cutoff}$ can sometimes be crucial because, as explained in \refsec{sec: theory_EF}, in certain cases one can find infinite friction for extremely small $\Gamma(\bm{r})$.
	We must emphasize, however, that {\bf  this divergence of friction is not an artifact.} In fact, this divergence in friction actually forces EF dynamics to recover Marcus's  theory of electrochemical charge transfer in the {\em nonadiabatic regime} for a one-dimensional quantum Brownian oscillator model\cite{Ouyang2016ele_transfer_EF_CME_compare}. And yet, however, the existence of an infinite frictional tensor must give one doubt about the overall applicability of EF dynamics. 
	To investigate the practical consequences of this divergence, we will now modify the original $T(z)$ model with the new parameters in \reftable{table: parameters_in_cut_off}; this substitution forces $T(z)$ to be much sharper than before. $x_1$ is kept at $0.2$. The modified parameters are plotted in \reffig{fig: surf_for_cutoff_anal}. With these new parameters and surfaces, we report relaxation rates from scattering simulations as a function of $\Gamma_{cutoff}$.  

\begin{figure}
		\centering
		\subfloat[]{\includegraphics[width=3in]{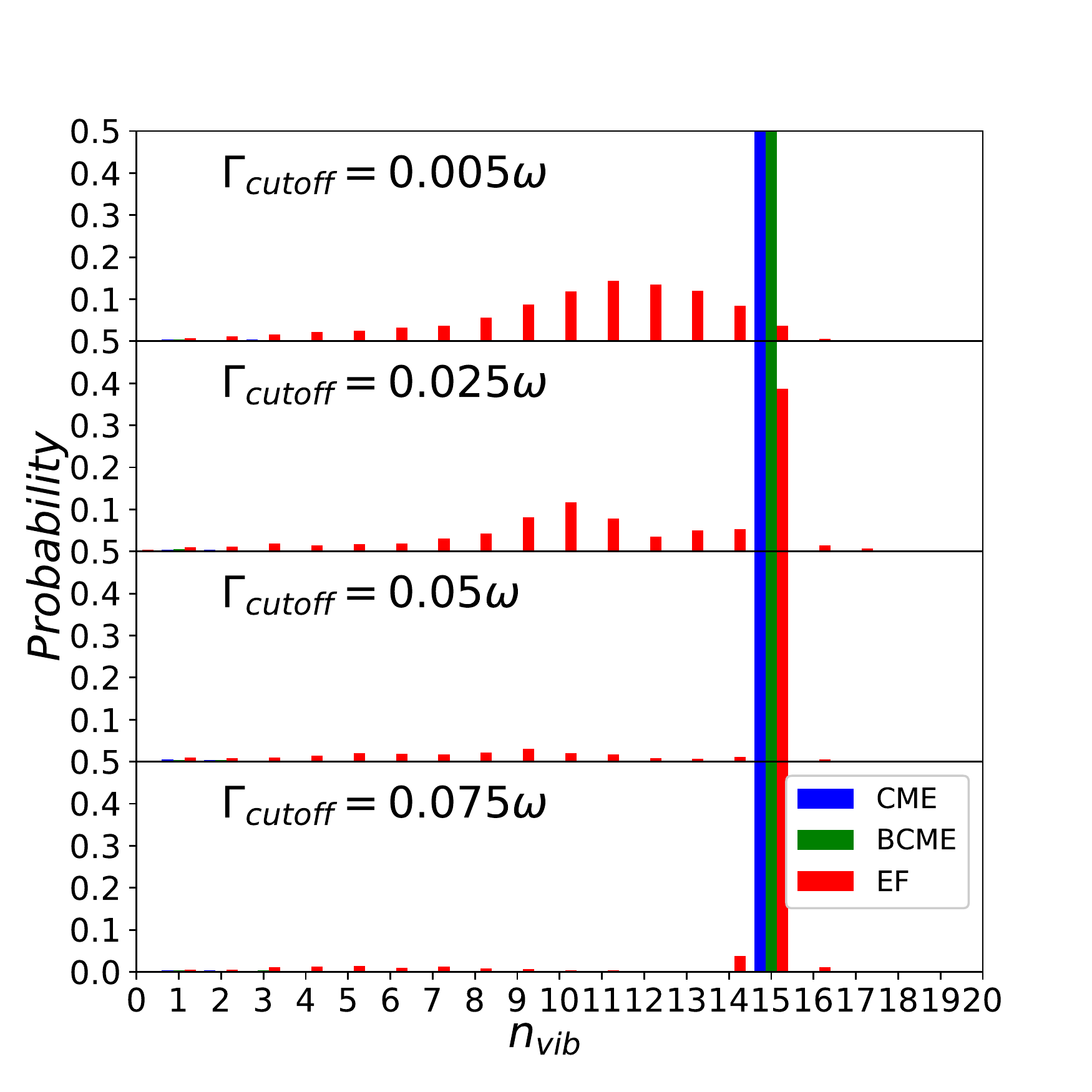}\label{fig: cutoff_anal}}
		\quad
		\subfloat[]{\includegraphics[width=3in]{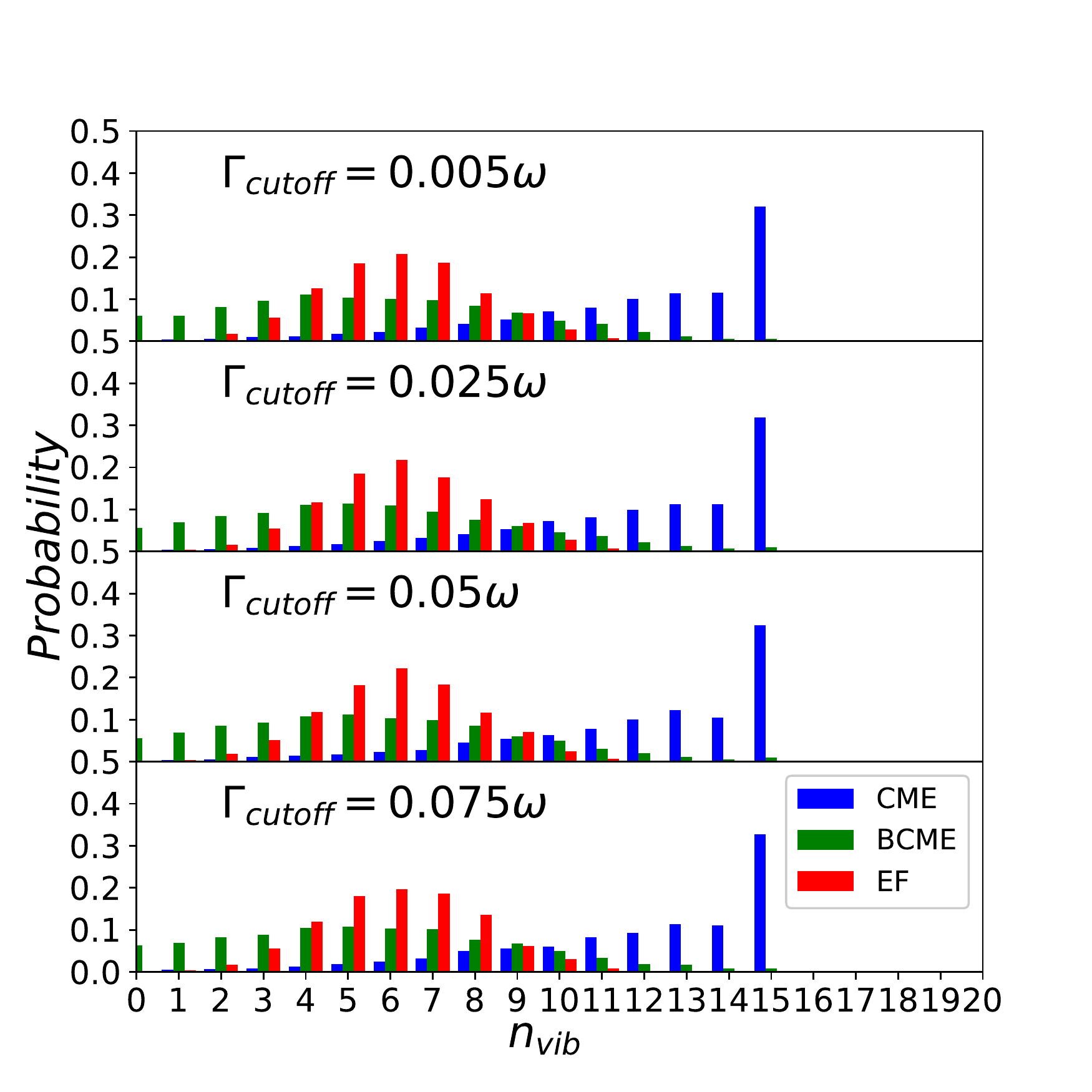}\label{fig: cutoff_anal_vib_surf}}
		\caption{ Vibrational distribution analysis for the scattered trajectories moving along the modified parameters in \reffig{fig: surf_for_cutoff_anal} (left) and the original surfaces in \reffig{fig: surf_for_vib_anal} (right). Here  $x_1 = 0.2$, $\Gamma_0 = 0.03$, $\avg{p_0} = 40$. For all $\Gamma_{cutoff}$ parameters used here, the cut-off region (where $\Gamma(\bm{r}) = \Gamma_{cutoff}$) is to the left of the surface crossing point. In the original model, the vibrational distributions from EF are consistent for different $\Gamma_{cutoff}$ parameters, but for the modified parameters, the distributions are quite sensitive to the artificial parameter $\Gamma_{cutoff}$. }
		\label{vib_anal_CUT}
	\end{figure}

	From the data in \reffig{fig: cutoff_anal}, we find that EF results are not equivalent for different values of $\Gamma_{cutoff}$. Indeed, for these parameters and such a small value of $\Gamma_0$, we find that the friction tensor is extremely large (nearly divergent). Thus, if $\Gamma_{cutoff}$ is very small, we find that EF dynamics can actually (and spuriously) predict more vibrational relaxation than BCME or CME dynamics. Luckily, this issue should not be important when $\Gamma(\bm{r})$ is not infinitesimal near a surface crossing region, as shown in \reffig{fig: cutoff_anal_vib_surf}. 

	Note that, except for \reffig{vib_anal_CUT}, all results reported in this paper using electronic friction can be considered reliable and converged with respect to $\Gamma_{cutoff}$ (see \reffig{vib_anal_G} and \reffig{vib_anal_V}). 
	\begin{table}[ht]
		\centering 
		\caption{different parameters in the cut-off analysis (compared to \reftable{table: parameters_in_vib})} 
		\begin{tabular}{c c c c } 
			\hline
			Parameter & Value(a.u.) & \ \ \ \  &Comment\\ 
			\hline 
			$c_g$ & 4& & parameter in \refeq{eq: gamma_shape}\\
			$z_g$ & -0.55 & & parameter in \refeq{eq: gamma_shape}\\
			\hline 
		\end{tabular}
		\label{table: parameters_in_cut_off} 
	\end{table} 
	
	Obviously, looking forward, the fact that BCME requires no such artificial parameter is a huge relative advantage. Unlike the case of EF, both BCME and CME dynamics propose simple smooth dynamics along diabats in regions with $\Gamma(\bm{r}) \rightarrow 0$.

\section{Conclusions and Future Directions}\label{sec: conclusion}

	In summary, we have investigated vibrational relaxation within a 2D scattering simulation where we  expect transient electron transfer for a variety of different approaches. Our conclusions are as follows:

	\begin{itemize}

		\item We find that the CME approach is unable to predict accurate vibrational relaxation probabilities. Whenever the metal-molecule coupling $\Gamma$ is large, CME dynamics along the simple diabatic curves are usually not accurate. In particular, the trajectory often misses the crossing region entirely.  These dynamics usually disagree with BCME and EF dynamics (even for large $\Gamma$).

	\item We find that EF dynamics give reasonable probabilities of vibrational relaxation (and thus agree with BCME dynamics) when $\Gamma(\bm{r})$ is reasonably large in the surface crossing region. However, there are clearly spurious effects when $\Gamma$ becomes too small.

	\item Overall, the BCME approach appears to give the most sensible data. By construction, this algorithm mostly agrees with the CME algorithm (in the limit of small $\Gamma$) and with the EF algorithm (in the limit of large $\Gamma$). The BCME approach tends to be more sensitive to electron-phonon couplings and the BCME approach usually results in more relaxation and a slightly wider distribution of vibrational quanta than do EF dynamics. 

	\end{itemize}
		
	Finally, perhaps the most surprising conclusion of this work is the prediction that there is a turnover in the rate of vibrational relaxation for scattering experiments as a function of $\Gamma$. According to \reffig{vib_anal_G}, we predict that the probability for vibrational relaxation peaks when $\Gamma$ is neither too small nor too large. This turnover feature is not found in condensed phase dynamics, where the rate of molecule-metal electron transfer is strictly increasing with the coupling parameter $\Gamma$: see Fig. 2 in Ref \onlinecite{Ouyang2016ele_transfer_EF_CME_compare}. It would be very interesting to identify a series of different metal substrates with varying degrees of metal-molecule coupling ($\Gamma$) from which this trend could be confirmed experimentally.
		
	Lastly, looking forward, we have two clear next steps. 
	First, given the simplicity	of the BCME approach (which ignores electronic coherences for a two-state problem), it will be very interesting to compare the BCME algorithm with IESH\cite{Shenvi2009IESH} (which includes coherences within the framework of a discretized metal). Such a comparison will tell us a great deal about when and why the BCME works/fails.
	Second, in order to apply the present dynamics to a real (and not model) system, it	will be essential to extract  (rather than conjecture) the relevant parameters from {\em ab initio} electronic structure calculations. This work is ongoing.
	
\section{Acknowledgements}
This material is based upon work supported by the (U.S.) Air Force Office of Scientific Research (USAFOSR) PECASE award under AFOSR Grant No. FA9950-13-1-0157. J.E.S. acknowledges a Camille and Henry Dreyfus Teacher Scholar Award and a David and Lucille Packard Fellowship.

\bibliographystyle{apsrev}

\begin{thebibliography}{0}
\expandafter\ifx\csname natexlab\endcsname\relax\def\natexlab#1{#1}\fi
\expandafter\ifx\csname bibnamefont\endcsname\relax
  \def\bibnamefont#1{#1}\fi
\expandafter\ifx\csname bibfnamefont\endcsname\relax
  \def\bibfnamefont#1{#1}\fi
\expandafter\ifx\csname citenamefont\endcsname\relax
  \def\citenamefont#1{#1}\fi
\expandafter\ifx\csname url\endcsname\relax
  \def\url#1{\texttt{#1}}\fi
\expandafter\ifx\csname urlprefix\endcsname\relax\def\urlprefix{URL }\fi
\providecommand{\bibinfo}[2]{#2}
\providecommand{\eprint}[2][]{\url{#2}}

\end{thebibliography}


\begin{thebibliography}{41}
\expandafter\ifx\csname natexlab\endcsname\relax\def\natexlab#1{#1}\fi
\expandafter\ifx\csname bibnamefont\endcsname\relax
  \def\bibnamefont#1{#1}\fi
\expandafter\ifx\csname bibfnamefont\endcsname\relax
  \def\bibfnamefont#1{#1}\fi
\expandafter\ifx\csname citenamefont\endcsname\relax
  \def\citenamefont#1{#1}\fi
\expandafter\ifx\csname url\endcsname\relax
  \def\url#1{\texttt{#1}}\fi
\expandafter\ifx\csname urlprefix\endcsname\relax\def\urlprefix{URL }\fi
\providecommand{\bibinfo}[2]{#2}
\providecommand{\eprint}[2][]{\url{#2}}

\bibitem[{\citenamefont{Wong and
  Rossky}(2002{\natexlab{a}})}]{Wong2002SolventDissipation}
\bibinfo{author}{\bibfnamefont{K.~F.} \bibnamefont{Wong}} \bibnamefont{and}
  \bibinfo{author}{\bibfnamefont{P.~J.} \bibnamefont{Rossky}},
  \bibinfo{journal}{The Journal of chemical physics}
  \textbf{\bibinfo{volume}{116}}, \bibinfo{pages}{8429}
  (\bibinfo{year}{2002}{\natexlab{a}}).

\bibitem[{\citenamefont{Wong and
  Rossky}(2002{\natexlab{b}})}]{Wong2002SolventDissipation2}
\bibinfo{author}{\bibfnamefont{K.~F.} \bibnamefont{Wong}} \bibnamefont{and}
  \bibinfo{author}{\bibfnamefont{P.~J.} \bibnamefont{Rossky}},
  \bibinfo{journal}{The Journal of chemical physics}
  \textbf{\bibinfo{volume}{116}}, \bibinfo{pages}{8418}
  (\bibinfo{year}{2002}{\natexlab{b}}).

\bibitem[{\citenamefont{Jailaubekov et~al.}(2013)\citenamefont{Jailaubekov,
  Willard, Tritsch, Chan, Sai, Gearba, Kaake, Williams, Leung, Rossky
  et~al.}}]{Jailaubekov2013PhotoExcitation}
\bibinfo{author}{\bibfnamefont{A.~E.} \bibnamefont{Jailaubekov}},
  \bibinfo{author}{\bibfnamefont{A.~P.} \bibnamefont{Willard}},
  \bibinfo{author}{\bibfnamefont{J.~R.} \bibnamefont{Tritsch}},
  \bibinfo{author}{\bibfnamefont{W.-L.} \bibnamefont{Chan}},
  \bibinfo{author}{\bibfnamefont{N.}~\bibnamefont{Sai}},
  \bibinfo{author}{\bibfnamefont{R.}~\bibnamefont{Gearba}},
  \bibinfo{author}{\bibfnamefont{L.~G.} \bibnamefont{Kaake}},
  \bibinfo{author}{\bibfnamefont{K.~J.} \bibnamefont{Williams}},
  \bibinfo{author}{\bibfnamefont{K.}~\bibnamefont{Leung}},
  \bibinfo{author}{\bibfnamefont{P.~J.} \bibnamefont{Rossky}},
  \bibnamefont{et~al.}, \bibinfo{journal}{Nature materials}
  \textbf{\bibinfo{volume}{12}}, \bibinfo{pages}{66} (\bibinfo{year}{2013}).

\bibitem[{\citenamefont{Nelson et~al.}(2014)\citenamefont{Nelson,
  Fernandez-Alberti, Roitberg, and
  Tretiak}}]{Nelson2014PhotoExcitationDynamicsRev}
\bibinfo{author}{\bibfnamefont{T.}~\bibnamefont{Nelson}},
  \bibinfo{author}{\bibfnamefont{S.}~\bibnamefont{Fernandez-Alberti}},
  \bibinfo{author}{\bibfnamefont{A.~E.} \bibnamefont{Roitberg}},
  \bibnamefont{and} \bibinfo{author}{\bibfnamefont{S.}~\bibnamefont{Tretiak}},
  \bibinfo{journal}{Accounts of chemical research}
  \textbf{\bibinfo{volume}{47}}, \bibinfo{pages}{1155} (\bibinfo{year}{2014}).

\bibitem[{\citenamefont{Schwartz et~al.}(1996)\citenamefont{Schwartz, Bittner,
  Prezhdo, and Rossky}}]{Schwartz1996CondensePhaseNonAdiab}
\bibinfo{author}{\bibfnamefont{B.~J.} \bibnamefont{Schwartz}},
  \bibinfo{author}{\bibfnamefont{E.~R.} \bibnamefont{Bittner}},
  \bibinfo{author}{\bibfnamefont{O.~V.} \bibnamefont{Prezhdo}},
  \bibnamefont{and} \bibinfo{author}{\bibfnamefont{P.~J.}
  \bibnamefont{Rossky}}, \bibinfo{journal}{The Journal of chemical physics}
  \textbf{\bibinfo{volume}{104}}, \bibinfo{pages}{5942} (\bibinfo{year}{1996}).

\bibitem[{\citenamefont{Schwerdtfeger et~al.}(2014)\citenamefont{Schwerdtfeger,
  Soudackov, and Hammes-Schiffer}}]{Schwerdtfeger2014SolutionNonadiab}
\bibinfo{author}{\bibfnamefont{C.~A.} \bibnamefont{Schwerdtfeger}},
  \bibinfo{author}{\bibfnamefont{A.~V.} \bibnamefont{Soudackov}},
  \bibnamefont{and}
  \bibinfo{author}{\bibfnamefont{S.}~\bibnamefont{Hammes-Schiffer}},
  \bibinfo{journal}{The Journal of chemical physics}
  \textbf{\bibinfo{volume}{140}}, \bibinfo{pages}{034113}
  (\bibinfo{year}{2014}).

\bibitem[{\citenamefont{Wodtke et~al.}(2003)\citenamefont{Wodtke, Huang, and
  Auerbach}}]{Wodtke2003NOLiF}
\bibinfo{author}{\bibfnamefont{A.~M.} \bibnamefont{Wodtke}},
  \bibinfo{author}{\bibfnamefont{Y.}~\bibnamefont{Huang}}, \bibnamefont{and}
  \bibinfo{author}{\bibfnamefont{D.~J.} \bibnamefont{Auerbach}},
  \bibinfo{journal}{The Journal of chemical physics}
  \textbf{\bibinfo{volume}{118}}, \bibinfo{pages}{8033} (\bibinfo{year}{2003}).

\bibitem[{\citenamefont{Huang et~al.}(2000)\citenamefont{Huang, Rettner,
  Auerbach, and Wodtke}}]{Huang2000NOAuexp}
\bibinfo{author}{\bibfnamefont{Y.}~\bibnamefont{Huang}},
  \bibinfo{author}{\bibfnamefont{C.~T.} \bibnamefont{Rettner}},
  \bibinfo{author}{\bibfnamefont{D.~J.} \bibnamefont{Auerbach}},
  \bibnamefont{and} \bibinfo{author}{\bibfnamefont{A.~M.}
  \bibnamefont{Wodtke}}, \bibinfo{journal}{Science}
  \textbf{\bibinfo{volume}{290}}, \bibinfo{pages}{111} (\bibinfo{year}{2000}).

\bibitem[{\citenamefont{White et~al.}(2005)\citenamefont{White, Chen, Matsiev,
  Auerbach, and Wodtke}}]{White2005NOAuExp_vibrelax}
\bibinfo{author}{\bibfnamefont{J.~D.} \bibnamefont{White}},
  \bibinfo{author}{\bibfnamefont{J.}~\bibnamefont{Chen}},
  \bibinfo{author}{\bibfnamefont{D.}~\bibnamefont{Matsiev}},
  \bibinfo{author}{\bibfnamefont{D.~J.} \bibnamefont{Auerbach}},
  \bibnamefont{and} \bibinfo{author}{\bibfnamefont{A.~M.}
  \bibnamefont{Wodtke}}, \bibinfo{journal}{Nature}
  \textbf{\bibinfo{volume}{433}}, \bibinfo{pages}{503} (\bibinfo{year}{2005}).

\bibitem[{\citenamefont{Wodtke* et~al.}(2004)\citenamefont{Wodtke*, Tully, and
  Auerbach}}]{Wodtke2004nonadiabaticRev}
\bibinfo{author}{\bibfnamefont{A.~M.} \bibnamefont{Wodtke*}},
  \bibinfo{author}{\bibfnamefont{J.~C.} \bibnamefont{Tully}}, \bibnamefont{and}
  \bibinfo{author}{\bibfnamefont{D.~J.} \bibnamefont{Auerbach}},
  \bibinfo{journal}{International Reviews in Physical Chemistry}
  \textbf{\bibinfo{volume}{23}}, \bibinfo{pages}{513} (\bibinfo{year}{2004}).

\bibitem[{\citenamefont{Bartels et~al.}(2011)\citenamefont{Bartels, Cooper,
  Auerbach, and Wodtke}}]{Bartels2011nonadiab_minirev}
\bibinfo{author}{\bibfnamefont{C.}~\bibnamefont{Bartels}},
  \bibinfo{author}{\bibfnamefont{R.}~\bibnamefont{Cooper}},
  \bibinfo{author}{\bibfnamefont{D.~J.} \bibnamefont{Auerbach}},
  \bibnamefont{and} \bibinfo{author}{\bibfnamefont{A.~M.}
  \bibnamefont{Wodtke}}, \bibinfo{journal}{Chemical Science}
  \textbf{\bibinfo{volume}{2}}, \bibinfo{pages}{1647} (\bibinfo{year}{2011}).

\bibitem[{\citenamefont{Wodtke}(2016)}]{Wodtke2016Rev}
\bibinfo{author}{\bibfnamefont{A.~M.} \bibnamefont{Wodtke}},
  \bibinfo{journal}{Chemical Society Reviews} \textbf{\bibinfo{volume}{45}},
  \bibinfo{pages}{3641} (\bibinfo{year}{2016}).

\bibitem[{\citenamefont{Bartels et~al.}(2014)\citenamefont{Bartels, Kr{\"u}ger,
  Auerbach, Wodtke, and Sch{\"a}fer}}]{Bartels2014NOAu_moreexp}
\bibinfo{author}{\bibfnamefont{N.}~\bibnamefont{Bartels}},
  \bibinfo{author}{\bibfnamefont{B.~C.} \bibnamefont{Kr{\"u}ger}},
  \bibinfo{author}{\bibfnamefont{D.~J.} \bibnamefont{Auerbach}},
  \bibinfo{author}{\bibfnamefont{A.~M.} \bibnamefont{Wodtke}},
  \bibnamefont{and}
  \bibinfo{author}{\bibfnamefont{T.}~\bibnamefont{Sch{\"a}fer}},
  \bibinfo{journal}{Angewandte Chemie International Edition}
  \textbf{\bibinfo{volume}{53}}, \bibinfo{pages}{13690} (\bibinfo{year}{2014}).

\bibitem[{\citenamefont{Katz et~al.}(2005)\citenamefont{Katz, Zeiri, and
  Kosloff}}]{Katz2005ModelForNOAu}
\bibinfo{author}{\bibfnamefont{G.}~\bibnamefont{Katz}},
  \bibinfo{author}{\bibfnamefont{Y.}~\bibnamefont{Zeiri}}, \bibnamefont{and}
  \bibinfo{author}{\bibfnamefont{R.}~\bibnamefont{Kosloff}},
  \bibinfo{journal}{The Journal of Physical Chemistry B}
  \textbf{\bibinfo{volume}{109}}, \bibinfo{pages}{18876}
  (\bibinfo{year}{2005}).

\bibitem[{\citenamefont{Galperin and
  Nitzan}(2015)}]{Galperin2015MoleculeMetalInterface}
\bibinfo{author}{\bibfnamefont{M.}~\bibnamefont{Galperin}} \bibnamefont{and}
  \bibinfo{author}{\bibfnamefont{A.}~\bibnamefont{Nitzan}},
  \bibinfo{journal}{The journal of physical chemistry letters}
  \textbf{\bibinfo{volume}{6}}, \bibinfo{pages}{4898} (\bibinfo{year}{2015}).

\bibitem[{\citenamefont{Head-Gordon and
  Tully}(1992)}]{HeadGordon1992COCutheory}
\bibinfo{author}{\bibfnamefont{M.}~\bibnamefont{Head-Gordon}} \bibnamefont{and}
  \bibinfo{author}{\bibfnamefont{J.~C.} \bibnamefont{Tully}},
  \bibinfo{journal}{The Journal of chemical physics}
  \textbf{\bibinfo{volume}{96}}, \bibinfo{pages}{3939} (\bibinfo{year}{1992}).

\bibitem[{\citenamefont{Rittmeyer et~al.}(2015)\citenamefont{Rittmeyer, Meyer,
  Juaristi, and Reuter}}]{Rittmeyer2015EF_application}
\bibinfo{author}{\bibfnamefont{S.~P.} \bibnamefont{Rittmeyer}},
  \bibinfo{author}{\bibfnamefont{J.}~\bibnamefont{Meyer}},
  \bibinfo{author}{\bibfnamefont{J.~I.} \bibnamefont{Juaristi}},
  \bibnamefont{and} \bibinfo{author}{\bibfnamefont{K.}~\bibnamefont{Reuter}},
  \bibinfo{journal}{Physical review letters} \textbf{\bibinfo{volume}{115}},
  \bibinfo{pages}{046102} (\bibinfo{year}{2015}).

\bibitem[{\citenamefont{Nitzan and
  Tully}(1983)}]{Nitzan1983EF_compare_to_quantum_pertubation}
\bibinfo{author}{\bibfnamefont{A.}~\bibnamefont{Nitzan}} \bibnamefont{and}
  \bibinfo{author}{\bibfnamefont{J.~C.} \bibnamefont{Tully}},
  \bibinfo{journal}{The Journal of Chemical Physics}
  \textbf{\bibinfo{volume}{78}}, \bibinfo{pages}{3959} (\bibinfo{year}{1983}).

\bibitem[{\citenamefont{Li and Wahnstr{\"o}m}(1992)}]{Li1992EFsimu}
\bibinfo{author}{\bibfnamefont{Y.}~\bibnamefont{Li}} \bibnamefont{and}
  \bibinfo{author}{\bibfnamefont{G.}~\bibnamefont{Wahnstr{\"o}m}},
  \bibinfo{journal}{Physical review letters} \textbf{\bibinfo{volume}{68}},
  \bibinfo{pages}{3444} (\bibinfo{year}{1992}).

\bibitem[{\citenamefont{Frischkorn and
  Wolf}(2006)}]{Frischkorn2006NonadiabaticReactionRev}
\bibinfo{author}{\bibfnamefont{C.}~\bibnamefont{Frischkorn}} \bibnamefont{and}
  \bibinfo{author}{\bibfnamefont{M.}~\bibnamefont{Wolf}},
  \bibinfo{journal}{Chemical reviews} \textbf{\bibinfo{volume}{106}},
  \bibinfo{pages}{4207} (\bibinfo{year}{2006}).

\bibitem[{\citenamefont{Misewich et~al.}(1992)\citenamefont{Misewich, Heinz,
  and Newns}}]{Misewich1992DIMET}
\bibinfo{author}{\bibfnamefont{J.}~\bibnamefont{Misewich}},
  \bibinfo{author}{\bibfnamefont{T.}~\bibnamefont{Heinz}}, \bibnamefont{and}
  \bibinfo{author}{\bibfnamefont{D.}~\bibnamefont{Newns}},
  \bibinfo{journal}{Physical review letters} \textbf{\bibinfo{volume}{68}},
  \bibinfo{pages}{3737} (\bibinfo{year}{1992}).

\bibitem[{\citenamefont{Bard}(2010)}]{Bard2010HeterogenousElectrodeReaction}
\bibinfo{author}{\bibfnamefont{A.~J.} \bibnamefont{Bard}},
  \bibinfo{journal}{Journal of the American Chemical Society}
  \textbf{\bibinfo{volume}{132}}, \bibinfo{pages}{7559} (\bibinfo{year}{2010}).

\bibitem[{\citenamefont{Shenvi et~al.}(2009{\natexlab{a}})\citenamefont{Shenvi,
  Roy, and Tully}}]{Shenvi2009IESH}
\bibinfo{author}{\bibfnamefont{N.}~\bibnamefont{Shenvi}},
  \bibinfo{author}{\bibfnamefont{S.}~\bibnamefont{Roy}}, \bibnamefont{and}
  \bibinfo{author}{\bibfnamefont{J.~C.} \bibnamefont{Tully}},
  \bibinfo{journal}{The Journal of chemical physics}
  \textbf{\bibinfo{volume}{130}}, \bibinfo{pages}{174107}
  (\bibinfo{year}{2009}{\natexlab{a}}).

\bibitem[{\citenamefont{Shenvi et~al.}(2009{\natexlab{b}})\citenamefont{Shenvi,
  Roy, and Tully}}]{Shenvi2009NOAusimu_sci}
\bibinfo{author}{\bibfnamefont{N.}~\bibnamefont{Shenvi}},
  \bibinfo{author}{\bibfnamefont{S.}~\bibnamefont{Roy}}, \bibnamefont{and}
  \bibinfo{author}{\bibfnamefont{J.~C.} \bibnamefont{Tully}},
  \bibinfo{journal}{Science} \textbf{\bibinfo{volume}{326}},
  \bibinfo{pages}{829} (\bibinfo{year}{2009}{\natexlab{b}}).

\bibitem[{\citenamefont{Monturet and
  Saalfrank}(2010)}]{Monturet2010NOAu_EF_simu}
\bibinfo{author}{\bibfnamefont{S.}~\bibnamefont{Monturet}} \bibnamefont{and}
  \bibinfo{author}{\bibfnamefont{P.}~\bibnamefont{Saalfrank}},
  \bibinfo{journal}{Physical Review B} \textbf{\bibinfo{volume}{82}},
  \bibinfo{pages}{075404} (\bibinfo{year}{2010}).

\bibitem[{\citenamefont{Li and Guo}(2002)}]{Li2002MCmodelforNOAu}
\bibinfo{author}{\bibfnamefont{S.}~\bibnamefont{Li}} \bibnamefont{and}
  \bibinfo{author}{\bibfnamefont{H.}~\bibnamefont{Guo}}, \bibinfo{journal}{The
  Journal of chemical physics} \textbf{\bibinfo{volume}{117}},
  \bibinfo{pages}{4499} (\bibinfo{year}{2002}).

\bibitem[{\citenamefont{Head-Gordon and
  Tully}(1995)}]{HeadGordon1995electronicfriction}
\bibinfo{author}{\bibfnamefont{M.}~\bibnamefont{Head-Gordon}} \bibnamefont{and}
  \bibinfo{author}{\bibfnamefont{J.~C.} \bibnamefont{Tully}},
  \bibinfo{journal}{The Journal of chemical physics}
  \textbf{\bibinfo{volume}{103}}, \bibinfo{pages}{10137}
  (\bibinfo{year}{1995}).

\bibitem[{\citenamefont{Brandbyge et~al.}(1995)\citenamefont{Brandbyge,
  Hedeg{\aa}rd, Heinz, Misewich, and Newns}}]{Brandbyge1995elefric}
\bibinfo{author}{\bibfnamefont{M.}~\bibnamefont{Brandbyge}},
  \bibinfo{author}{\bibfnamefont{P.}~\bibnamefont{Hedeg{\aa}rd}},
  \bibinfo{author}{\bibfnamefont{T.}~\bibnamefont{Heinz}},
  \bibinfo{author}{\bibfnamefont{J.}~\bibnamefont{Misewich}}, \bibnamefont{and}
  \bibinfo{author}{\bibfnamefont{D.}~\bibnamefont{Newns}},
  \bibinfo{journal}{Physical Review B} \textbf{\bibinfo{volume}{52}},
  \bibinfo{pages}{6042} (\bibinfo{year}{1995}).

\bibitem[{\citenamefont{Dou and Subotnik}(2017)}]{Dou2017nonconstGammaEF}
\bibinfo{author}{\bibfnamefont{W.}~\bibnamefont{Dou}} \bibnamefont{and}
  \bibinfo{author}{\bibfnamefont{J.~E.} \bibnamefont{Subotnik}},
  \bibinfo{journal}{The Journal of Chemical Physics}
  \textbf{\bibinfo{volume}{146}}, \bibinfo{pages}{092304}
  (\bibinfo{year}{2017}).

\bibitem[{\citenamefont{Dou et~al.}(2016)\citenamefont{Dou, Nitzan, and
  Subotnik}}]{Dou2016hesterisis}
\bibinfo{author}{\bibfnamefont{W.}~\bibnamefont{Dou}},
  \bibinfo{author}{\bibfnamefont{A.}~\bibnamefont{Nitzan}}, \bibnamefont{and}
  \bibinfo{author}{\bibfnamefont{J.~E.} \bibnamefont{Subotnik}},
  \bibinfo{journal}{The Journal of chemical physics}
  \textbf{\bibinfo{volume}{144}}, \bibinfo{pages}{074109}
  (\bibinfo{year}{2016}).

\bibitem[{\citenamefont{Maurer et~al.}(2016)\citenamefont{Maurer, Askerka,
  Batista, and Tully}}]{Maurer2016abinit_friction_tensor}
\bibinfo{author}{\bibfnamefont{R.~J.} \bibnamefont{Maurer}},
  \bibinfo{author}{\bibfnamefont{M.}~\bibnamefont{Askerka}},
  \bibinfo{author}{\bibfnamefont{V.~S.} \bibnamefont{Batista}},
  \bibnamefont{and} \bibinfo{author}{\bibfnamefont{J.~C.} \bibnamefont{Tully}},
  \bibinfo{journal}{Physical Review B} \textbf{\bibinfo{volume}{94}},
  \bibinfo{pages}{115432} (\bibinfo{year}{2016}).

\bibitem[{\citenamefont{Elste et~al.}(2008)\citenamefont{Elste, Weick, Timm,
  and von Oppen}}]{Elste2008CME_original}
\bibinfo{author}{\bibfnamefont{F.}~\bibnamefont{Elste}},
  \bibinfo{author}{\bibfnamefont{G.}~\bibnamefont{Weick}},
  \bibinfo{author}{\bibfnamefont{C.}~\bibnamefont{Timm}}, \bibnamefont{and}
  \bibinfo{author}{\bibfnamefont{F.}~\bibnamefont{von Oppen}},
  \bibinfo{journal}{Applied Physics A: Materials Science \& Processing}
  \textbf{\bibinfo{volume}{93}}, \bibinfo{pages}{345} (\bibinfo{year}{2008}).

\bibitem[{\citenamefont{Dou et~al.}(2015)\citenamefont{Dou, Nitzan, and
  Subotnik}}]{Dou2015CME}
\bibinfo{author}{\bibfnamefont{W.}~\bibnamefont{Dou}},
  \bibinfo{author}{\bibfnamefont{A.}~\bibnamefont{Nitzan}}, \bibnamefont{and}
  \bibinfo{author}{\bibfnamefont{J.~E.} \bibnamefont{Subotnik}},
  \bibinfo{journal}{The Journal of chemical physics}
  \textbf{\bibinfo{volume}{143}}, \bibinfo{pages}{054103}
  (\bibinfo{year}{2015}).

\bibitem[{\citenamefont{Dou and Subotnik}(2016{\natexlab{a}})}]{Dou2016bCME}
\bibinfo{author}{\bibfnamefont{W.}~\bibnamefont{Dou}} \bibnamefont{and}
  \bibinfo{author}{\bibfnamefont{J.~E.} \bibnamefont{Subotnik}},
  \bibinfo{journal}{The Journal of chemical physics}
  \textbf{\bibinfo{volume}{144}}, \bibinfo{pages}{024116}
  (\bibinfo{year}{2016}{\natexlab{a}}).

\bibitem[{\citenamefont{Dou and
  Subotnik}(2016{\natexlab{b}})}]{Dou2016manybodyEF}
\bibinfo{author}{\bibfnamefont{W.}~\bibnamefont{Dou}} \bibnamefont{and}
  \bibinfo{author}{\bibfnamefont{J.~E.} \bibnamefont{Subotnik}},
  \bibinfo{journal}{The Journal of Chemical Physics}
  \textbf{\bibinfo{volume}{145}}, \bibinfo{pages}{054102}
  (\bibinfo{year}{2016}{\natexlab{b}}).

\bibitem[{\citenamefont{Newns}(1986)}]{Newns1986NOsurfacemodel}
\bibinfo{author}{\bibfnamefont{D.}~\bibnamefont{Newns}},
  \bibinfo{journal}{Surface science} \textbf{\bibinfo{volume}{171}},
  \bibinfo{pages}{600} (\bibinfo{year}{1986}).

\bibitem[{\citenamefont{Ouyang et~al.}(2016)\citenamefont{Ouyang, Dou, Jain,
  and Subotnik}}]{Ouyang2016ele_transfer_EF_CME_compare}
\bibinfo{author}{\bibfnamefont{W.}~\bibnamefont{Ouyang}},
  \bibinfo{author}{\bibfnamefont{W.}~\bibnamefont{Dou}},
  \bibinfo{author}{\bibfnamefont{A.}~\bibnamefont{Jain}}, \bibnamefont{and}
  \bibinfo{author}{\bibfnamefont{J.~E.} \bibnamefont{Subotnik}},
  \bibinfo{journal}{Journal of Chemical Theory and Computation}
  \textbf{\bibinfo{volume}{12}}, \bibinfo{pages}{4178} (\bibinfo{year}{2016}).

\bibitem[{\citenamefont{Schmickler and
  Santos}(2010)}]{Schmickler2010InterfacialElectrochemistry}
\bibinfo{author}{\bibfnamefont{W.}~\bibnamefont{Schmickler}} \bibnamefont{and}
  \bibinfo{author}{\bibfnamefont{E.}~\bibnamefont{Santos}},
  \emph{\bibinfo{title}{Interfacial electrochemistry}}
  (\bibinfo{publisher}{Springer Science \& Business Media},
  \bibinfo{year}{2010}).

\bibitem[{\citenamefont{LaRue et~al.}(2011{\natexlab{a}})\citenamefont{LaRue,
  Schäfer, Matsiev, Velarde, Nahler, Auerbach, and
  Wodtke}}]{Larue2011NOAu_kinetic_energy_distribution}
\bibinfo{author}{\bibfnamefont{J.~L.} \bibnamefont{LaRue}},
  \bibinfo{author}{\bibfnamefont{T.}~\bibnamefont{Schäfer}},
  \bibinfo{author}{\bibfnamefont{D.}~\bibnamefont{Matsiev}},
  \bibinfo{author}{\bibfnamefont{L.}~\bibnamefont{Velarde}},
  \bibinfo{author}{\bibfnamefont{N.~H.} \bibnamefont{Nahler}},
  \bibinfo{author}{\bibfnamefont{D.~J.} \bibnamefont{Auerbach}},
  \bibnamefont{and} \bibinfo{author}{\bibfnamefont{A.~M.}
  \bibnamefont{Wodtke}}, \bibinfo{journal}{The Journal of Physical Chemistry A}
  \textbf{\bibinfo{volume}{115}}, \bibinfo{pages}{14306}
  (\bibinfo{year}{2011}{\natexlab{a}}).

\bibitem[{\citenamefont{LaRue et~al.}(2011{\natexlab{b}})\citenamefont{LaRue,
  Sch{\"a}fer, Matsiev, Velarde, Nahler, Auerbach, and
  Wodtke}}]{Larue2011NOAu_kinetic_energy_distribution2}
\bibinfo{author}{\bibfnamefont{J.}~\bibnamefont{LaRue}},
  \bibinfo{author}{\bibfnamefont{T.}~\bibnamefont{Sch{\"a}fer}},
  \bibinfo{author}{\bibfnamefont{D.}~\bibnamefont{Matsiev}},
  \bibinfo{author}{\bibfnamefont{L.}~\bibnamefont{Velarde}},
  \bibinfo{author}{\bibfnamefont{N.~H.} \bibnamefont{Nahler}},
  \bibinfo{author}{\bibfnamefont{D.~J.} \bibnamefont{Auerbach}},
  \bibnamefont{and} \bibinfo{author}{\bibfnamefont{A.~M.}
  \bibnamefont{Wodtke}}, \bibinfo{journal}{Physical Chemistry Chemical Physics}
  \textbf{\bibinfo{volume}{13}}, \bibinfo{pages}{97}
  (\bibinfo{year}{2011}{\natexlab{b}}).

\bibitem[{\citenamefont{Shenvi et~al.}(2006)\citenamefont{Shenvi, Roy,
  Parandekar, and Tully}}]{Shenvi2006NOAuTheory_vibrelax_model}
\bibinfo{author}{\bibfnamefont{N.}~\bibnamefont{Shenvi}},
  \bibinfo{author}{\bibfnamefont{S.}~\bibnamefont{Roy}},
  \bibinfo{author}{\bibfnamefont{P.}~\bibnamefont{Parandekar}},
  \bibnamefont{and} \bibinfo{author}{\bibfnamefont{J.}~\bibnamefont{Tully}},
  \bibinfo{journal}{The Journal of chemical physics}
  \textbf{\bibinfo{volume}{125}}, \bibinfo{pages}{154703}
  (\bibinfo{year}{2006}).

\end{thebibliography}

\end{document}